\documentclass[acmsmall]{acmart}
\renewcommand\footnotetextcopyrightpermission[1]{} % removes footnote with conference information in first column
\AtBeginDocument{%
  }

\usepackage{algpseudocode}
\usepackage[linesnumbered,ruled,vlined]{algorithm2e}
\usepackage{pifont}
\usepackage{graphicx}
\usepackage{textcomp}
\usepackage{xcolor}
\usepackage{pgf}
\usepackage{tikz}
\usetikzlibrary{shapes.geometric, arrows.meta, positioning, shadows, fit, calc, backgrounds}
\usepackage{color}
\usepackage{stfloats}
\usepackage{multirow}
\usepackage{diagbox}
\usepackage{booktabs}
\usepackage{arydshln}
\usepackage{lipsum}
\usepackage{wrapfig}
\usepackage{array}
\usepackage{booktabs}
\usepackage{tabularx} 
\usepackage{ragged2e}  
\usepackage{enumitem}
\usepackage[normalem]{ulem}
\renewcommand{\arraystretch}{1.2} % 可选：调行距
\usepackage{hyperref}
\usepackage{makecell}
\usepackage{array} % 提供更多的列格式选项

\usepackage{xcolor}
\usepackage{soul}
\usepackage{amssymb}
\usepackage{comment}
\usepackage{wrapfig}
\usepackage[normalem]{ulem}

\newcommand{\squishlist}{
   \begin{list}{$\bullet$}
    { \setlength{\itemsep}{0pt}      \setlength{\parsep}{0pt}
      \setlength{\topsep}{3pt}       \setlength{\partopsep}{0pt}
      \setlength{\listparindent}{-2pt}
      \setlength{\itemindent}{-5pt}
      \setlength{\leftmargin}{1em} \setlength{\labelwidth}{0em}
      \setlength{\labelsep}{0.5em} } }

\newcommand{\squishend}{
    \end{list}  }

\newcommand{\todo}[1]{{\color{red}\sf\bfseries \hl{[TODO]} #1}}

\newcommand{\RemoveRebuttal}[1]{{}}

\setuldepth{cat}
\newcommand{\note}[1]{{\color{magenta}$\square$}}

\usepackage{tikz}
\usetikzlibrary{shapes,positioning}
\usepackage{hyperref}

\newcounter{insight}
\newcommand{\insight}{%
  \refstepcounter{insight}% 为了和 \label / \ref 联动
  \tikz[baseline=(char.base)]{
    \node[shape=circle,draw,inner sep=0.99pt,fill=black,text=white] (char) {\scriptsize\bfseries \theinsight};
  }\hspace{0.5em}\textbf%
}

\newcommand{\insightref}[1]{%
  \hyperref[#1]{%
    \tikz[baseline=(char.base)]{
      \node[shape=circle,draw,inner sep=0.99pt,fill=black,text=white] (char) {\scriptsize\bfseries \ref*{#1}};
    }%
  }%
}
\begin{document}

% \title{LLM Inference on GPUs: A Comprehensive Experimental Review, Characterization, and Optimization Insights}
\title{
%Characterizing LLM Inference on GPUs: A Comprehensive Study from Micro-architectural Bounds to System-level Insights
A Systematic Characterization of LLM Inference on GPUs
}


\author{Haonan Wang}
\email{wanghaonan241@mails.ucas.ac.cn}
\orcid{0009-0007-1495-0604}

\author{Xuxin Xiao}
\email{xiaoxuxin25e@ict.ac.cn}
\orcid{0009-0003-3449-4467}

\author{Mingyu Yan}
\authornote{Corresponding author.}
\email{yanmingyu@ict.ac.cn}
\orcid{0000-0002-6915-955X}

\author{Zhuoyuan Zhu}
\email{zhuzhuoyuan24s@ict.ac.cn}
\orcid{0009-0009-9350-0808}

\author{Dengke Han}
\email{handengke21s@ict.ac.cn}
\orcid{0000-0003-0641-5779}

\author{Duo Wang}
\email{wangduo18z@ict.ac.cn}
\orcid{0000-0001-9363-7796}

\author{Wenming Li}
\email{liwenming@ict.ac.cn}
\orcid{0000-0003-4069-2251}

\author{Xiaochun Ye}
\email{yexiaochun@ict.ac.cn}
\orcid{0000-0003-4598-1685}
\affiliation{%
  \institution{State Key Lab of Processors, Institute of Computing Technology, Chinese Academy of Sciences
  %; University of Chinese Academy of Sciences
  }
  \city{Beijing}
  \country{China}
  \postcode{100190}
}

\author{Cunchen Hu}
\email{hucc12@chinatelecom.cn}
\orcid{0009-0004-0605-3792}
\affiliation{%
  \institution{China Telecom Cloud Computing Research Institute}
  \city{Beijing}
  \country{China}
}

\author{Hongyang Chen}
\email{hongyang@zhejianglab.com}
\orcid{0000-0002-7626-0162}
\affiliation{%
  \institution{Zhejiang Lab}
  \city{Hangzhou}
  \country{China}
}

\author{Guangyu Sun}
\email{gsun@pku.edu.cn}
\orcid{0000-0002-7315-6589}
\affiliation{%
  \institution{Peking University}
  \city{Beijing}
  \country{China}
}

\renewcommand{\shortauthors}{H. Wang et al.}


\begin{abstract}

This work presents a systematic characterization of  Large Language Model (LLM) inference to address fragmented understanding. Through comprehensive experiments, we establish a four-dimensional analytical framework: (1) Two-Phase Heterogeneity Observation; (2) Microarchitectural Root Cause Analysis; (3) System Scaling Principles; and (4) Emerging Paradigm Boundaries. Our investigation progresses systematically from observation to foresight: identifying performance phenomena, revealing hardware causes, validating system behavior, and exploring new paradigms. This study not only consolidates a reliable empirical foundation for existing research but also provides new discoveries and practical optimization guidance for LLM inference.
%the co-design of future LLM architectures and inference systems.

%This work presents a systematic characterization of LLM inference on GPUs to address the fragmented nature of existing studies. Through a unified experimental framework, we establish a comprehensive understanding organized around four analytical dimensions: (1) Phenomenon Observation: Establishing Two-Phase Heterogeneity; (2) Root Cause Explanation: Uncovering Microarchitectural Etiology; (3) Scaling Effect Analysis: Validating System-Level Governing Principles; and (4) Boundary Redefinition: Exploring Paradigm Shifts.

%Our investigation progresses systematically from observation to foresight: we first establish what the core performance phenomena are, then determine why they occur at the hardware level, subsequently validate how these principles govern behavior in realistic systems, and finally explore where performance boundaries are being redefined by emerging paradigms. This study not only consolidates a reliable empirical foundation for existing research but also provides new discoveries and practical guidance for the co-design of future LLM architectures and inference systems.

\end{abstract}

\maketitle

\section{Introduction}

The inference of Large Language Models (LLMs) has emerged as a critical workload spanning from data centers to edge devices~\cite{zhao2023survey,hadi2023large,yi2023edgemoe,Tian2025CLONECL,Namboori2025DistributedIO,chen2025characterizing,zheng2025review}. As LLM deployment expands across diverse applications—from real-time conversational systems~\cite{kaddour2023challenges,liu2024deepseekv3,thoppilan2022lamda} to complex retrieval-augmented generation~\cite{gao2023retrieval,huang2024survey,fan2024survey,lewis2020retrieval} pipelines—optimizing inference efficiency has become increasingly vital~\cite{yuan2024llm,zhou2024survey,zhen2025taming,li2024large,park2025survey,li2024llm,qin2024mooncake,lee2024infinigen}. Unlike conventional deep learning workloads, LLM inference exhibits a distinctive two-phase execution pattern: the parallelizable Prefill phase and the sequential Decode phase~\cite{Zhong2024DistServe,Kamath2024PODAttentionUF}. This inherent computational heterogeneity generates conflicting hardware demands that dictate system performance across latency, throughput, and energy efficiency metrics. 

Although the compute-bound nature of Prefill and memory-bound characteristics of Decode have gained broad recognition, current research predominantly treats this dichotomy as a static assumption rather than examining it through dynamic quantitative analysis. The growing complexity of model architectures—particularly the shift toward Mixture-of-Experts (MoE)~\cite{Fedus2021SwitchTS,cai2025survey,shazeer2017outrageously} paradigms—coupled with expanding deployment scenarios encompassing edge computing and sophisticated Retrieval-Augmented Generation (RAG) workflows, has further complicated the interplay between hardware constraints and software characteristics. This landscape creates a pressing need for a systematic characterization that can not only validate established understandings with rigorous microarchitectural evidence but also reveal fundamental relationships between execution patterns, memory hierarchy behavior, and energy consumption profiles.

Existing characterization studies exhibit significant fragmentation across research domains. Device- and energy-focused analyses profile platform-specific latency and power patterns~\cite{Patel2024CharacterizingPM,benazir2025profiling,moghadampanah2025energy,fernandez2025energy,Kubwimana2025EdgeReasoningCR}, while quantization research examines precision-performance trade-offs~\cite{shi2025systematic}. Kernel- and compiler-level investigations dissect attention/GEMM operations and optimization strategies~\cite{Dao2022FlashAttention,carmona2025characterization,Vellaisamy2025CharacterizingAO,zhang2024llmcompass}, and system-oriented studies explore serving, scheduling policies, and cache management~\cite{xu2025characterizing,Zhong2024DistServe,Stojkovic2024DynamoLLMDL,Siavashi2025PriorityAwarePS,Lazuka2024LLMPilotCA,Hu2024CharacterizationOL,ren2025characterizing,wang2025kvcache}. Despite this breadth, current work remains siloed—lacking the cross-layer, component-integrated perspective needed for holistic LLM inference characterization.

To bridge this gap, our research introduces a unified experimental framework that systematically synthesizes existing knowledge while exploring new analytical directions. This integrated approach enables us to provide architectural-level verification and refinement of established conclusions while yielding new insights in previously underexplored domains. Our systematic investigation progresses through four interconnected analytical dimensions. \uline{(1) Two-Phase Heterogeneity:} We establish the fundamental dichotomy between Prefill and Decode phases through cross-scenario measurements, defining phase heterogeneity as a core paradigm for LLM inference. \uline{(2) Microarchitectural Root Causes:} Using Roofline modeling and stall analysis, we identify the hardware origins of phase divergence through operator characteristics and data reuse patterns.
\uline{(3) Scaling Behavior:} We validate phase-aware parallelization strategies across multi-GPU systems, and shows how the bottlenecks of Prefill and Decode phases are amplified at the edge device. \uline{(4) Emerging Paradigms:} We characterize sparse activation tradeoffs in MoEs models and bottleneck migration in RAG workflows, informing next-generation system design.

Our findings establish a systematic framework for understanding LLM inference. These insights provide both theoretical foundations for future model-system co-design and practical optimization guidances for real-world deployment across diverse hardware environments.

\section{Background}
\label{sec:background}

\subsection{LLM Inference}
\label{subsec:inference_process}

\begin{figure}[!t]

    \centering
    \includegraphics[width=1.0\textwidth]{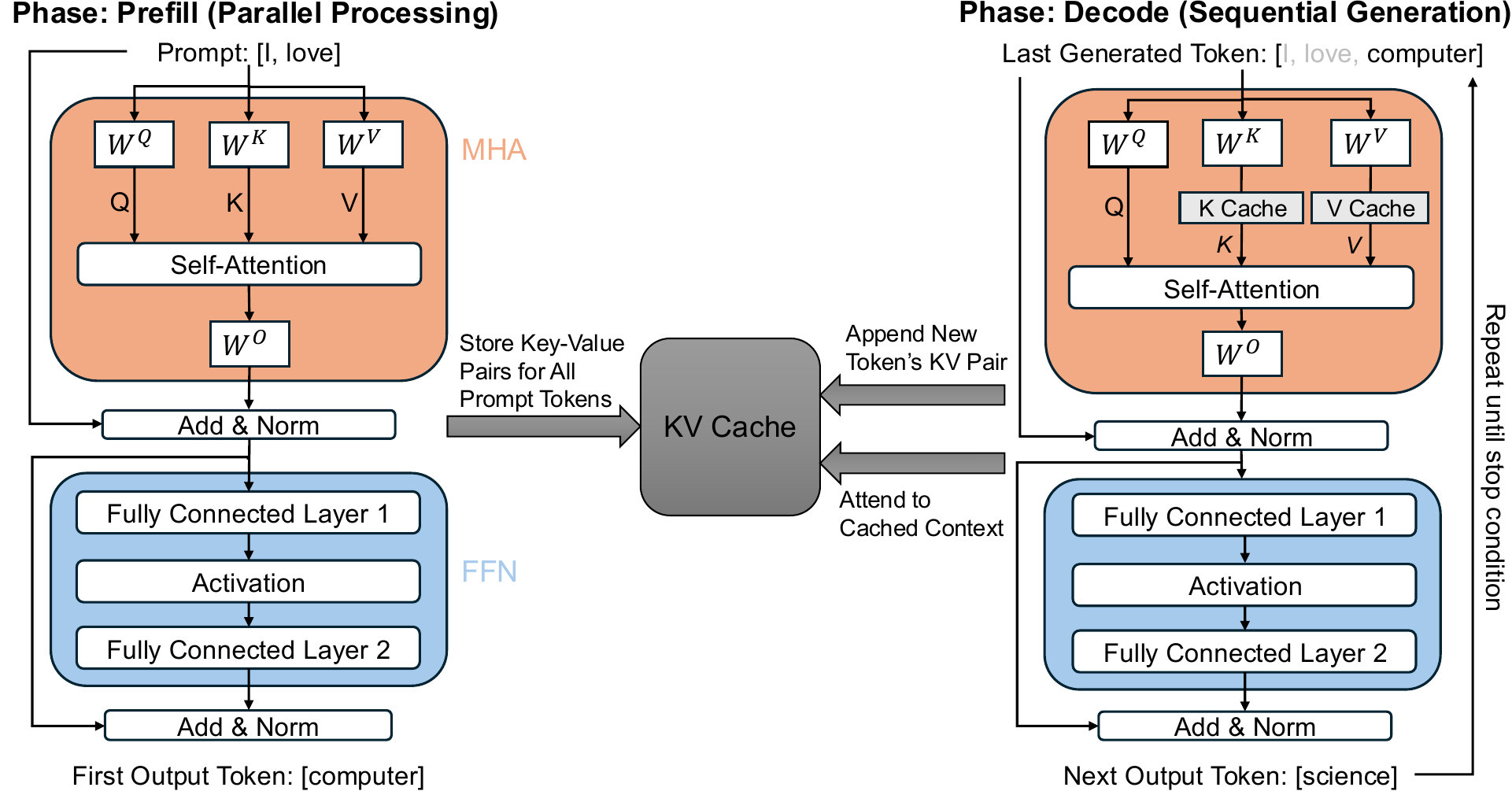}
    \vspace{-15pt}
    \caption{Illustration of the LLM inference process, comprising the Prefill and Decode phases.}
    \label{fig:background_pipeline}
        \vspace{-15pt}
\end{figure}

We focus on decoder-only Transformer architectures (e.g., Llama-3, Qwen2.5). As shown in Fig.~\ref{fig:background_pipeline}, the inference process naturally decomposes into two distinct phases. \uline{(1) Prefill Phase.} The model processes the entire input prompt in parallel. The self-attention mechanism computes the Key and Value tensors for all prompt tokens and stores them in the \textit{KV Cache}. This parallel computation establishes the initial context required for subsequent generation. \uline{(2) Decode Phase.} The model generates output tokens autoregressively, one step at a time. Each step processes only the single newly generated token. To maintain context, the attention mechanism uses the \textit{KV Cache} to retrieve the Key and Value tensors of all preceding tokens, computing attention scores against the full history before appending the current step's KV pairs to the cache.

%In our implementation, all attention computations are realized using FlashAttention-style fused kernels; unless otherwise stated, we refer to this fused attention kernel as \texttt{Attn.core} and use this name in our later kernel-level analysis.

%We focus on decoder-only Transformer architectures (e.g., Llama-3, Qwen2.5). As illustrated in Fig.~\ref{fig:background_pipeline}, the inference process naturally decomposes into two distinct phases:

%\textbf{Prefill Phase.} The model processes the complete input prompt of length $L_{\text{in}}$ in parallel. During this phase, the self-attention mechanism computes the Key and Value matrices for all prompt tokens and stores them in the \textit{KV Cache}. This parallel computation establishes the initial context required for subsequent generation.

%\textbf{Decode Phase.} The model generates output tokens autoregressively, one step at a time. In each step, the model processes only the single newly generated token. To maintain context, the attention mechanism utilizes the \textit{KV Cache} to retrieve the Key and Value tensors of all preceding tokens, computing attention scores against the full history before appending the current step's KV pairs to the cache.

%In our implementation, all attention computations are realized using FlashAttention-style fused kernels; unless otherwise stated, we refer to this fused attention kernel as \texttt{Attn.core} and use this name in our later kernel-level analysis.

\subsection{Emerging Paradigms: MoE and RAG}

%Beyond conventional dense architectures, two emerging paradigms are reshaping the computational characteristics of LLM inference. 

While dense Transformer architectures have dominated early LLM development, two emerging paradigms are fundamentally reshaping the computational characteristics of LLM inference.

\begin{figure}[!t]
        \vspace{10pt}

    \centering
    \includegraphics[width=1.0\textwidth]{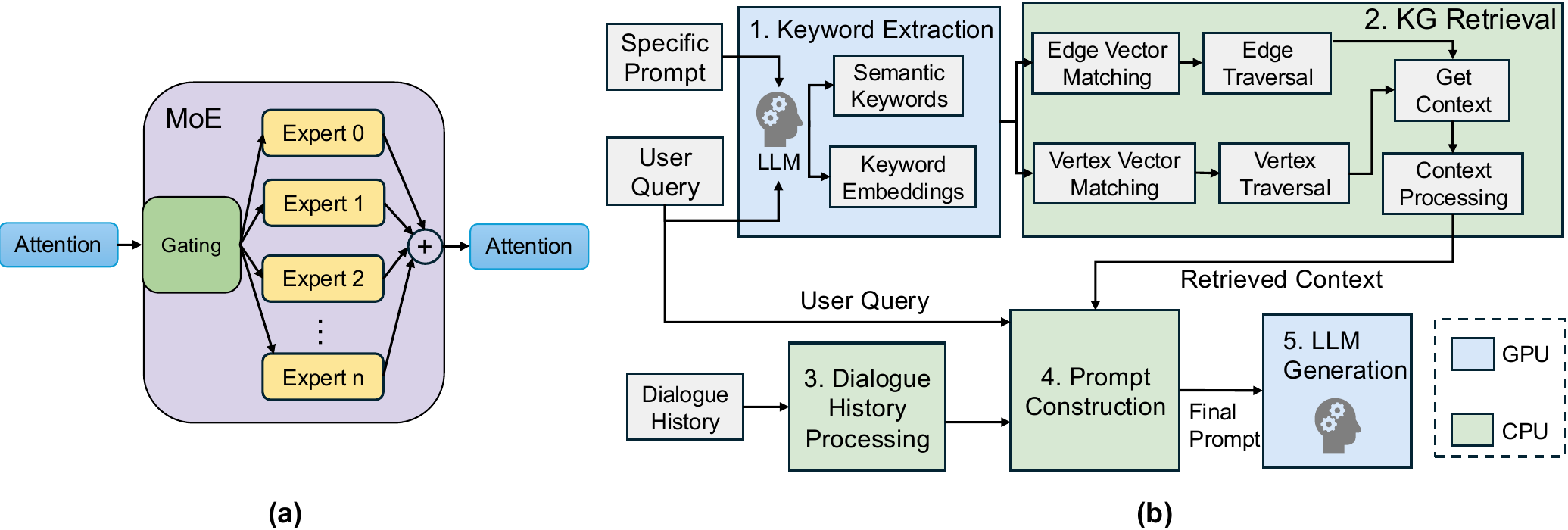}
        \vspace{-15pt}
   \caption{Overview of Emerging Paradigms: (a) MoE architecture; (b) RAG workflow.}
    \label{fig:MoE_RAG_overview}
        \vspace{-15pt}

\end{figure}

\uline{Mixture-of-Experts.} 
MoE represents a fundamental architectural shift from dense computation to sparse activation. By replacing dense feed-forward networks with specialized expert layers, MoE architectures achieve unprecedented decoupling of model capacity from computational cost. As illustrated in Fig.~\ref{fig:MoE_RAG_overview}(a), a routing network dynamically selects only the top-$k$ experts per token, enabling massive parameter counts while activating minimal computational paths. This sparse activation dramatically reduces FLOPs per token but introduces distinct system challenges: significant routing overhead emerges during sequential Decode, while expert partitioning creates fragmented memory access patterns, establishing new phase-dependent optimization frontiers.

\uline{Retrieval-Augmented Generation.} 
RAG transforms LLM inference by integrating explicit knowledge retrieval components. These systems create a heterogeneous CPU-GPU pipeline where external knowledge retrieval precedes LLM execution, fundamentally redistributing system bottlenecks. As shown in Fig.~\ref{fig:MoE_RAG_overview}(b), performance constraints shift from GPU computational throughput to CPU-side retrieval latency and memory bandwidth saturation. Furthermore, the integration of retrieved knowledge substantially expands effective context length during the Prefill phase, introducing additional computational complexity and memory demands that reshape the entire inference profile.

The RAG workflow comprises five distinct stages. \uline{Keyword Extraction}: LLM processes the input query to extract semantic keywords and generate embeddings. \uline{Knowledge Graph (KG) Retrieval}: Hybrid vector and graph-based searches retrieve relevant contextual information. \uline{Dialogue History Processing}: Management and preparation of previous conversation turns. \uline{Prompt Construction}: Integration of retrieved context, dialogue history, and current query into the final prompt. \uline{LLM Generation}: Standard inference execution on the augmented prompt.

The KG Retrieval stage further decomposes into six functional sub-stages. \uline{Edge/Vertex Vector Matching}: Compute vector similarities to identify relevant graph entities. \uline{Edge/Vertex Traversal}: Navigate graph structure to expand retrieval scope. \uline{Get Context}: Acquire textual content associated with retrieved entities. \uline{Context Processing}: Preprocess and prepare retrieved text for integration. The matching sub-stages handle similarity computation, while the traversal and processing sub-stages manage graph navigation and text preparation, respectively.

\section{Characterization Methodology}
\label{sec:methodology}

%In this section, we first present the methodology underlying our characterization. To support this methodology, we then outline our experimental setup, including the hardware platforms, inference framework, model selection, and workload construction. Finally, we describe our profiling and measurement approaches.

%In this section, we first introduce our characterization methodology, then describe the experimental setup covering hardware platforms, inference framework, model selection, and workload construction, and finally present our profiling and measurement approaches.

This section details our characterization methodology, experimental setup, and profiling approaches.

\subsection{Methodology}

Our characterization methodology is designed to systematically unravel the performance landscape of LLM inference, moving from symptomatic observation to prescriptive insight. The four steps form a causal chain: we first \uline{establish what} the core performance phenomena are, then \uline{determine why} they occur at the hardware level, subsequently \uline{validate how} these principles govern behavior in realistic systems, and finally \uline{explore where} the performance boundaries are being redefined by emerging paradigms. These steps are implemented as follows.

\uline{(1) Phenomenon Observation: Establishing the "What" – Two-Phase Heterogeneity.}

\squishlist

\item Objective: Identify the fundamental macroscopic performance characteristics.

\item Action: We systematically measure latency distribution, throughput, and energy across diverse application scenarios of LLM inference.

\item Outcome: This step establishes the core empirical fact of a fundamental dichotomy between the Prefill and Decode phases, documenting their distinct bottleneck profiles, dynamic migration, and predictable behaviors. It answers what the system's performance symptoms are.

\squishend

\uline{(2) Root Cause Explanation: Uncovering the "Why" – Microarchitectural Etiology.}

\squishlist

\item Objective: Diagnose the root causes of the two-phase performance dichotomy.

\item Action: We employ Roofline modeling to quantify arithmetic intensity, warp stall analysis to pinpoint pipeline bottlenecks, and memory behaviors profiling to assess memory locality.

\item Outcome: This step connects macroscopic symptoms to microarchitectural root causes, revealing that Prefill is inherently compute-bound due to high-intensity GEMMs, while Decode is memory-bound due to low-intensity, KV-cache-dominated data access.

\squishend

\uline{(3) Scaling Effect Analysis: Validating the "How" – System-Level Governing Principles.}

\squishlist

\item Objective: Test how phase-aware principles generalize to scaled and constrained environments.

\item Action: We extend the analysis to multi-GPU setups (evaluating Tensor and Pipeline Parallelism) and edge devices (profiling under resource constraints).

\item Outcome: This step demonstrates how the root causes dictate system design. It validates that the compute-bound Prefill scales with Tensor Parallelism, while the memory-bound Decode is better served by Pipeline Parallelism, and shows how these bottlenecks are amplified at the edge. 

\squishend

\uline{(4) Boundary Redefinition: Exploring the "Where" – Paradigm Shifts.}

\squishlist

\item Objective: Investigate how new architectural and workflow innovations reshape the established performance boundaries.

\item Action: We characterize sparse MoE models and RAG workflows, quantifying their new efficiency trade-offs and bottleneck migrations.

\item Outcome: This step looks forward to where the landscape is evolving. It shows how MoE decouples total parameters from active computation, and how RAG shifts the bottleneck from GPU to CPU, thereby redefining the optimization frontier for next-generation systems.

\squishend

This structured progression from observation to explanation, validation, and foresight ensures our characterization is not merely descriptive, but fundamentally explanatory and prescriptive.

\subsection{Experimental Setup}
\label{sec:experimental_setup}

To implement our characterization methodology, we established a comprehensive experimental framework comprising the platforms, models, workloads, and profiling methods detailed below.

\textbf{Platforms.} We employ two platforms spanning the compute spectrum: a high-performance server with two Intel Xeon Platinum 8350C CPUs and four NVIDIA A100 80GB SXM GPUs (interconnected via NVLink in pairs and PCIe across pairs); and an NVIDIA Jetson AGX Orin for edge-device characterization. All experiments use vLLM v0.9.2~\cite{kwon2023efficient} with BF16 precision.

%Our evaluation leverages two distinct platforms to capture performance characteristics across the compute spectrum. The primary platform is a high-performance server equipped with two Intel Xeon Platinum 8350C CPUs and four NVIDIA A100 80GB GPUs. These GPUs are interconnected in two groups via NVLink for high-bandwidth intra-group communication and PCIe for inter-group connectivity. This setup facilitates our analysis of multi-GPU parallelization strategies (Step 3). To characterize performance on resource-constrained devices, we additionally employ an NVIDIA Jetson AGX Orin developer kit, representing a typical edge AI platform. On the software side, we use vLLM~\cite{kwon2023efficient} version 0.9.2 as our inference engine, chosen for its widespread adoption and production-ready optimizations. All GPU experiments are conducted using the bfloat16 (BF16) precision format.  

\begin{table}[!t]
\vspace{-10pt}
\centering
\caption{Information of evaluated LLMs.}
\label{tab:llm_info}
\vspace{-10pt}
\small
\renewcommand\arraystretch{1.2}
\setlength\tabcolsep{2pt}%
\resizebox{0.9\textwidth}{!}{
\begin{tabular}{|c|c|c|c|c|c|c|}
\hline
Model & Params (B) & Layers & Hidden Dim. & Attn. Heads (Q/KV) & Arch. & Experts \\ \hline
Llama-3-8B-Instruct        & 8            & 32 & 4096 & 32 / 8 & Dense & --              \\ \hline
Qwen2.5-7B-Instruct        & 7            & 28 & 3584 & 28 / 4 & Dense & --              \\ \hline
Qwen2.5-32B-Instruct       & 32           & 64 & 5120 & 40 / 8 & Dense & --              \\ \hline
Qwen3-30B-A3B-Instruct-2507 & 30.5 / 3.3$^\dagger$ & 48 & 2048 & 32 / 4 & MoE   & 128 (top-8)    \\ \hline
\end{tabular}
}
\scriptsize

\noindent $^\dagger$Total / activated parameters per token, as reported by the official model card of Qwen3-30B-A3B-Instruct-2507.
\vspace{-15pt}

\end{table}

\textbf{Models.} Our evaluation covers four instruction-tuned LLMs (see Table~\ref{tab:llm_info}): the dense models Llama-3-8B-Instruct~\cite{dubey2024llama3} and Qwen2.5-7B/32B-Instruct~\cite{qwen2_5}, which share architectural similarity but differ in scale; and the sparse model Qwen3-30B-A3B-Instruct~\cite{qwen3-paper}, an MoE model activating 3.3B of its 30.5B total parameters per token. This selection provides a comprehensive testbed covering parameter scales from 7B to 32B across both dense and MoE architectures.

%We evaluate four open-source, instruction-tuned LLMs spanning two architectural paradigms including Llama~3~\cite{dubey2024llama3} and Qwen2.5/3~\cite{qwen2_5,qwen3-paper}, as detailed in Table~\ref{tab:llm_info}. For dense architectures, we select Llama-3-8B-Instruct~\cite{dubey2024llama3} and two Qwen2.5 models (7B/32B-Instruct)~\cite{qwen2_5}. These decoder-only Transformer models share similar architectures but vary significantly in scale, enabling analysis of model size effects. For sparse architectures, we include Qwen3-30B-A3B-Instruct-2507~\cite{qwen3-paper}, a Mixture-of-Experts model with 30.5B total parameters but only 3.3B activated parameters per token through its top-8 expert routing mechanism. This selection provides a comprehensive testbed covering parameter scales from 7B to 32B across both dense and MoE architectures.

\begin{table}[!t]

    \centering
    \caption{Summary of evaluation workloads.}
    \label{tab:workloads}
    \vspace{-10pt}
    \renewcommand\arraystretch{1.1}
    \setlength\tabcolsep{4pt}
    \small
    \resizebox{0.76\linewidth}{!}{
    \begin{tabular}{|l|l|c|c|c|}
        \hline
        Scenario & Dataset & Input length & Output length & \#Prompts \\
        \hline
        Chat       & ShareGPT-V3 (\texttt{sharegpt})        & variable (from dataset) & 128  & 16 \\ \hline
        Summary    & Sonnet (\texttt{sonnet})               & 8192                     & 256  & 8  \\ \hline
        Translate  & Synthetic (\texttt{random})            & 512                      & 512  & 16 \\ \hline
        Code       & InstructCoder (HF, \texttt{hf})        & variable (from dataset) & 2048 & 8  \\ \hline
        Story      & Synthetic (\texttt{random})            & 256                      & 3072 & 4  \\
        \hline
    \end{tabular}
    }
    \vspace{-15pt}
\end{table}

\textbf{Workloads.} We evaluate five representative workloads covering common LLM application scenarios, as configured in Table~\ref{tab:workloads}. These workloads are strategically chosen to elicit the two-phase behavior central to our analysis: 
1) Chat (ShareGPT-V3), interactive dialogue with variable-length inputs;
2) Summary (Sonnet corpus), long-context processing with 8K input tokens;
3) Translation (Synthetic), balanced input-output task;
4) Code Generation (InstructCoder): long-output scenario;
5) Story Writing (Synthetic): creative writing with extended generation.

All experiments use vLLM's offline inference benchmark with fixed request batches to isolate intrinsic execution behavior from arrival dynamics. The synthetic dataset enables precise control over prompt lengths, while real-world datasets provide realistic usage patterns.

\textbf{Profiling Methods.} We employ NVIDIA Nsight Systems/Compute for GPU analysis and Intel PCM for CPU microarchitecture characterization in RAG workflows. 
%The inference path is instrumented with NVTX markers for precise performance attribution to model components, focusing on the top-5 kernels accounting for >90\% (Prefill) and >85\% (Decode) of execution time. 
Experiments run in eager execution mode with high-frequency power sampling via nvidia-smi for energy measurements. All configurations are repeatedly executed and averaged to ensure statistical reliability.

\section{Phenomenon Observation — Establishing Two-phase Heterogeneity}
\label{sec:macro_level_characterization}

We begin our characterization by establishing the fundamental macroscopic performance phenomena of LLM inference. Through systematic measurements, we objectively document the intrinsic dichotomy between Prefill and Decode phases—two execution phases that exhibit fundamentally distinct performance signatures. Our analysis reveals that these phases differ radically in their GPU resource utilization patterns and energy efficiency characteristics. Furthermore, we observe dynamic bottleneck migration between phases as input/output lengths and model scales vary, alongside identifying strong predictable relationships: Prefill latency demonstrates a linear dependence on the number of uncached tokens, while total energy consumption scales linearly with output length. This phase establishes the essential ``symptom profile'' of LLM inference performance, providing the empirical foundation for subsequent root-cause investigation.

\subsection{Resource Utilization Divergence Between Prefill and Decode Phases}
\label{sec:two_phase_overview}

%As illustrated in Fig.~\ref{fig:macro_prefill_decode_breakdown}(a), the latency distribution between Prefill and Decode phases varies substantially across the five representative inference scenarios we evaluated, strongly correlating with application characteristics. In interactive tasks—such as Chat, Code, Story, and Translation—which typically involve short inputs and long outputs, the Decode phase accounts for 70\%–90\% of total latency. In contrast, for long-context tasks like document summarization (Summary), the one-time computational overhead of the Prefill phase constitutes a dominant portion of overall latency.
%
%These observations lead to a key insight: \insight{\label{ins:scenario_bottleneck}%The phase bottleneck of LLM inference is highly scenario-dependent: Decode dominates latency in interactive applications, whereas Prefill becomes the primary overhead in long-context processing tasks.} 

%Section~\ref{sec:dynamic_performance_behavior} will further analyze how variations in input/output length drive the dominance switching between Prefill and Decode.

\begin{figure}[!t]
\vspace{-10pt}

\centering
\includegraphics[width=1\textwidth]{figures/macro-level/macro_level_utilization_and_trend.pdf}
\vspace{-25pt}
\caption{(a) Overview of SM utilization and DRAM throughput across Prefill and Decode in the Chat scenario. (b) Transition of latency dominance from Decode to Prefill phase under increasing input length and fixed
output (128 tokens), measured on Qwen2.5-7B with a single GPU.}
\label{fig:macro_prefill_decode_breakdown}
\vspace{-10pt}
\end{figure}

We observe pronounced differences in how the two phases utilize GPU resources. As shown in Fig.~\ref{fig:macro_prefill_decode_breakdown}(a), Prefill exhibits significantly higher Streaming Multiprocessor (SM) utilization than Decode, indicating its stronger dependence on computational throughput. Conversely, Decode demonstrates notably higher memory bandwidth utilization, reflecting its greater demand for data access. This fundamental disparity in resource utilization patterns reveals the intrinsic difference in performance bottlenecks between the two phases.
These observations lead to a key insight: \insight{\label{ins:phase_resource_bottleneck}%
The Prefill and Decode phases are bottlenecked by different resources: Prefill by computational throughput and Decode by memory bandwidth.}

%To uncover the microarchitectural origins of this divergence, our subsequent analysis in Section~\ref{sec:micro_level_characterization} employs quantitative Roofline analysis.

\subsection{Dynamic Performance Behavior of Prefill and Decode Phases}
\label{sec:dynamic_performance_behavior}

LLM inference exhibits inherently dynamic performance characteristics, shaped by the complex interplay of workload properties (e.g., input/output lengths), model scales, and system configurations (e.g., batch size). To systematically deconstruct this behavior, we analyze its manifestations across three critical dimensions: first, we examine how latency bottlenecks migrate between phases and principal operators under varying conditions; second, we characterize the dynamic trade-off between throughput and latency as a function of batching strategy; and third, we investigate the predictability of Prefill latency, establishing its deterministic relationship with workload.

\subsubsection{Bottleneck Migration}
\label{sec:bottleneck_migration}

LLM inference is characterized by a dynamic bottleneck migration, where the dominant performance constraint shifts between execution phases and core operators. This subsection systematically unpacks how critical factors—namely input/output length and model scale—govern these transitions, revealing the context-dependent nature of inference performance.

%This subsection investigates the phenomenon of bottleneck migration in LLM inference, wherein the dominant performance constraint shifts between execution phases (Prefill vs. Decode) and operator types (e.g., FFN vs. Attention) under varying conditions. We systematically analyze how critical factors—including input/output length and model scale—drive these transitions, revealing the dynamic and context-dependent nature of inference bottlenecks.

% \begin{figure}[!t]
%     \centering
%     \includegraphics[width=0.9\textwidth]{figures/macro-level/prompt_length_trends.pdf}
%     \vspace{-10pt}
% \caption{Transition of latency dominance from Decode to Prefill phase under increasing input length and fixed output (128 tokens), measured on Qwen2.5-7B with a single GPU.}
%     \label{fig:macro_prompt_length_trends}
% \end{figure}

\uline{Phase-Level Bottleneck Migration.} As illustrated in Fig.~\ref{fig:macro_prefill_decode_breakdown}(b) for Qwen2.5-7B with a fixed output of 128 tokens $L_{out}$, the Decode phase accounts for most of the latency at short input lengths $L_{in}$. However, as input length grows, the Prefill latency share increases progressively, surpassing Decode after a critical point (around 2k tokens) to become the dominant bottleneck. Notably, the per-step Decode latency remains largely stable, independent of input length. This empirical analysis across models and workloads reveals a consistent pattern:
\insight{\label{ins:latency_crossover}%
A clear crossover point exists in the latency breakdown between Prefill and Decode as input length increases: Decode dominates at shorter inputs (forming Decode-Dominant regions), while Prefill becomes the bottleneck beyond a critical length (creating Prefill-Dominant regions).}
This crossover originates in the fundamental complexity differences between the two phases: Prefill performs one-time self-attention over the entire input, with quadratic complexity $O(L_{in}^2)$, whereas Decode reuses the KV cache and computes incrementally per token, leading to approximately linear complexity $O(L_{out})$ and stable per-step cost.

\begin{figure}[!t]
\vspace{-10pt}

    \centering
    \includegraphics[width=1\textwidth]{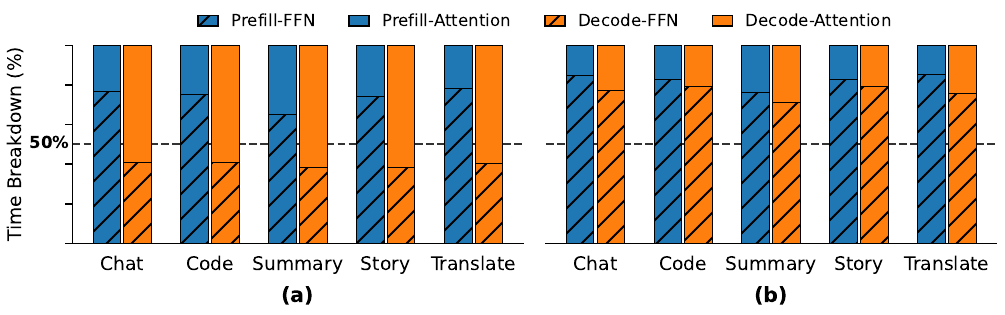}
    \vspace{-25pt}
    \caption{Operator-level bottleneck migration: transition between FFN-dominated and Attention-dominated latency as context length varies: (a) Llama-3-8B. (b) Qwen2.5-32B.}
    \label{fig:macro_ffn_attention_switch_comparison}
    \vspace{-10pt}
\end{figure}

\uline{Operator-Level Bottleneck Migration.}
Within LLM architectures, the FFN and Attention modules collectively account for the majority of inference latency, representing 94\% to 96\% of the total execution time across phases and scenarios. This predominance reveals that:
\insight{\label{ins:ffn_attention_dominant}%
The vast majority of inference latency is occupied by the FFN and Attention operators.}
We therefore investigate how the dominant operator shifts between FFN and Attention under varying conditions. As shown in Fig.~\ref{fig:macro_ffn_attention_switch_comparison}, measurements across typical context lengths (1k–8k) reveal consistent trends. For extremely long contexts (16k–32k+), prior studies~\cite{nawrot2025sparse,jiang2407minference} have established that Attention emerges as the primary bottleneck in both phases, owing to its rapidly escalating computational and memory overhead. In summary:
\insight{\label{ins:ffn_attention_switch}%
The bottleneck operator in LLM inference shifts between FFN and Attention as model scale and context length change.}

\begin{table}[!t]
\centering
\caption{Dominant operator types under different phases, context lengths, and model scales.}
\label{tab:phase_context_model_bottleneck_en}
\vspace{-10pt}
\small
\resizebox{0.8\textwidth}{!}{
\begin{tabular}{|l|l|c|c|}
\hline
Phase & Context Length ($n$) & Small Model (8B) & Large Model (32B) \\ \hline
\multirow{2}{*}{Prefill} 
& Typical context (1K-8K) & FFN  & FFN  \\ \cline{2-4}
& Extremely long context (>16K-32K+) & Attention  & Attention  \\
\hline
\multirow{2}{*}{Decode} 
& Typical context (1K-8K) & Attention & FFN  \\ \cline{2-4}
& Extremely long context (>16K-32K+) & Attention  & Attention  \\
\hline
\end{tabular}
}
\vspace{-10pt}
\end{table}

Table~\ref{tab:phase_context_model_bottleneck_en} details this switching pattern. The underlying mechanism lies in the differing computational complexities of FFN and Attention with respect to context length $n$. In the \textit{Prefill phase}, FFN scales as $O(n)$ while Attention scales as $O(n^2)$. For typical contexts (1K–8K), the quadratic term of Attention is not yet dominant, so the substantial fixed cost of FFN prevails. In extremely long contexts, however, Attention’s quadratic cost surpasses FFN, making it the primary bottleneck. In the \textit{Decode phase}, FFN cost per step is roughly constant $O(1)$, while Attention grows linearly $O(n)$ due to full KV-cache traversal. Thus, under typical contexts, large models (e.g., 32B) are FFN-dominated due to high fixed costs, while small models (e.g., 8B) are Attention-dominated. As context length extends further, Attention’s linear growth eventually exceeds FFN’s fixed cost, causing both model scales to become Attention-dominated in extremely long contexts.

\subsubsection{Throughput-Latency Trade-off}
\label{sec:throughput_latency_tradeoff}

We systematically characterize the fundamental yet conflicting relationship between throughput and latency in LLM inference, focusing on how batch size serves as the critical control parameter for balancing this trade-off.

\begin{figure}[!t]
\vspace{-10pt}

\centering
\includegraphics[width=1\textwidth]{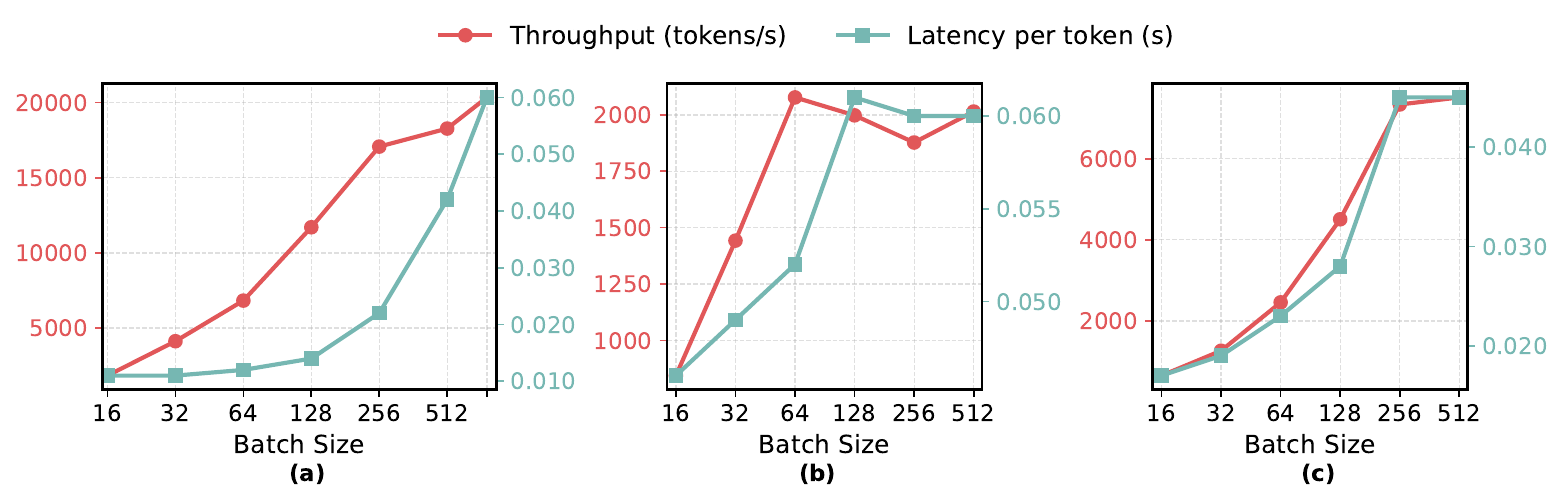}
\vspace{-25pt}
\caption{Characterizing the throughput versus latency trade-off under varying batch sizes for models (a) Qwen2.5-7B, (b) Qwen2.5-32B, and (c) Qwen3-30B-A3B.}
\label{fig:macro_throughput_vs_latency}
\vspace{-15pt}
\end{figure}

As illustrated in Fig.~\ref{fig:macro_throughput_vs_latency}, we characterize the throughput and latency trends under varying batch sizes across three representative models. Our analysis reveals a consistent pattern: \insight{\label{ins:throughput_latency_tradeoff}%
A dynamic throughput–latency trade-off exists: as batch size increases, system throughput (tokens/s) rises initially but exhibits diminishing returns, while average decoding latency grows monotonically, creating a fundamental performance conflict.}
This behavior stems from two competing effects: larger batches improve GPU utilization and amortize fixed overheads (e.g., kernel launching and scheduling), thereby boosting throughput. However, they also introduce increased queuing delays—since requests within a batch must complete jointly and decoding proceeds token-by-token, inter-step dependencies prolong average waiting time. Beyond a critical batch size, throughput gains plateau while latency penalties accelerate, leading to a \emph{latency trap}.

\subsubsection{Predictability of Prefill Latency}
\label{sec:predictability_of_prefill_latency}

The predictability of Prefill phase latency (Job Completion Time, JCT) is crucial for understanding system behavior and constructing effective scheduling models. Unlike the autoregressive Decode phase, which involves sequential dependencies and sampling variability, the computational workload of Prefill is determined solely by the number of input tokens, suggesting inherent deterministic behavior. Our experimental results confirm this hypothesis:
\insight{\label{ins:prefill_linear_uncached}%
JCT of Prefill phase exhibits a strong linear relationship with the number of uncached tokens, enabling highly accurate latency prediction.}

\begin{figure}[!t]
\centering
\includegraphics[width=1\textwidth]{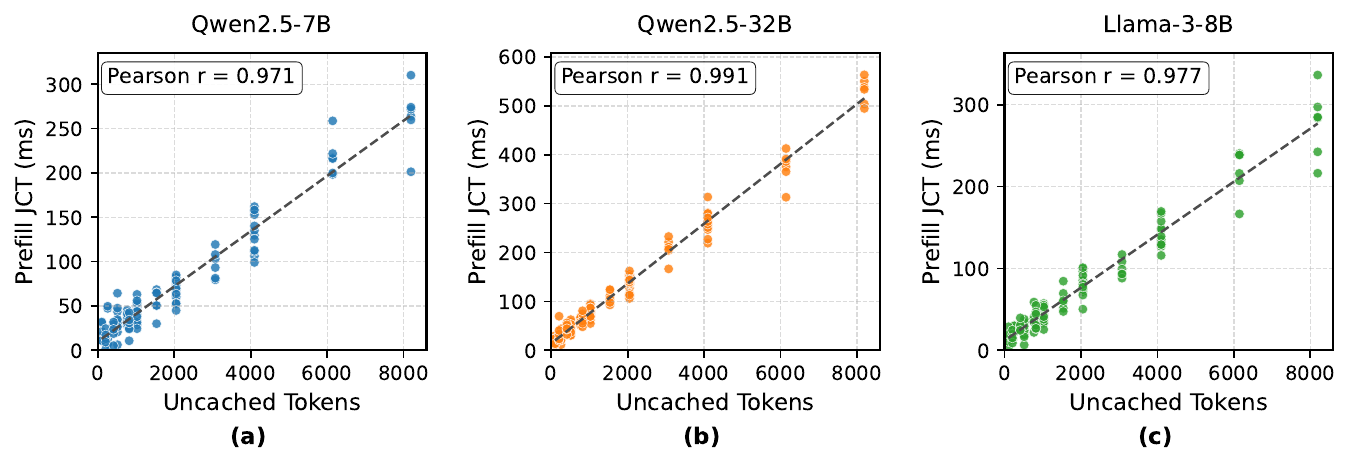}
\vspace{-25pt}
\caption{Prefill's JCT shows strong linear dependence on number of uncached tokens across models with high correlation and slope indicating per-token cost: (a) Qwen2.5-7B, (b) Qwen2.5-32B, and (c) Llama-3-8B.}
\label{fig:macro_prefill_predictability}
\vspace{-10pt}
\end{figure}

We systematically evaluate Prefill's JCT across three representative models. Input lengths covered short to long contexts \( n_{\text{input}} \in \{1\text{,}024,\; 2\text{,}048,\; 4\text{,}096,\; 8\text{,}192\} \), with KV cache hit ratios varied as \( r \in \{0\%, 25\%, 50\%, 75\%, 90\%\} \). For each \( (n_{\text{input}}, r) \) combination, we execute prefill-only inference—generating exactly one output token—multiple times, recording the Prefill's JCT and the number of uncached tokens, defined as: $n_{\text{uncached}} = n_{\text{input}} - n_{\text{cached}}$.
As shown in Fig.~\ref{fig:macro_prefill_predictability}, Prefill's JCT exhibits a strong linear correlation with \( n_{\text{uncached}} \) across all three models. The Pearson correlation coefficients are: \( c = 0.971 \) for Qwen2.5-7B, \( c = 0.991 \) for Qwen2.5-32B,  \( c = 0.977 \) for Llama-3-8B. 
The relationship is well-approximated by the linear model: $\text{JCT} \approx a + b \times n_{\text{uncached}}$, where the coefficient \( b \) reflects the average computational cost per uncached token, measured as approximately 0.0312~ms/token (Qwen2.5-7B), 0.0612~ms/token (Qwen2.5-32B), and 0.0324~ms/token (Llama-3-8B). Notably, the unit cost increases with model scale, indicating higher computational demand for larger models. A marginally weaker correlation observed at shorter input lengths (\( \sim1\text{K} \)) is due to the relatively larger impact of fixed overheads and measurement jitter in such regimes. 

%Overall, Prefill latency demonstrates a highly predictable linear dependence on the number of uncached tokens. This behavior originates from the computational structure of the Prefill phase: each uncached token necessitates full computation through the Attention and FFN layers, whereas cached tokens reuse previously computed KV results. 

%Hence, the value of \( n_{\text{uncached}} \) directly dictates the actual computational load, and under stable GPU execution conditions, Prefill JCT scales approximately linearly with \( n_{\text{uncached}} \).

\subsection{Energy Analysis}
\label{sec:energy_analysis}

Energy efficiency has become a critical concern in LLM inference, especially for large-scale deployments where power consumption directly impacts operational costs and environmental sustainability. 
%Unlike traditional deep learning models, LLM inference exhibits a unique two-phase execution pattern that significantly influences its energy characteristics.  
This subsection presents a comprehensive analysis of energy consumption patterns across different phases, output lengths, and model scales, with the goal of establishing predictable energy models.
We conduct energy measurements in a single A100 GPU environment. Our measurement methodology employs a precise power sampling approach. Before each inference task, we record GPU idle power for 10 seconds and compute the average as $P_{\text{baseline}}$.
During inference execution, we sample instantaneous power $P_{\text{sample}}(t)$ at 100ms intervals using \texttt{nvidia-smi --query-gpu=power.draw}.
Finally, we compute net energy consumption by integrating baseline-adjusted power over the task duration: $
E_{\text{net}} = \int_{\text{start}}^{\text{end}} \max\left(P_{\text{sample}}(t) - P_{\text{baseline}}, 0\right) dt$.

\begin{figure}[!t]
\vspace{-10pt}

\centering
\includegraphics[width=1\textwidth]{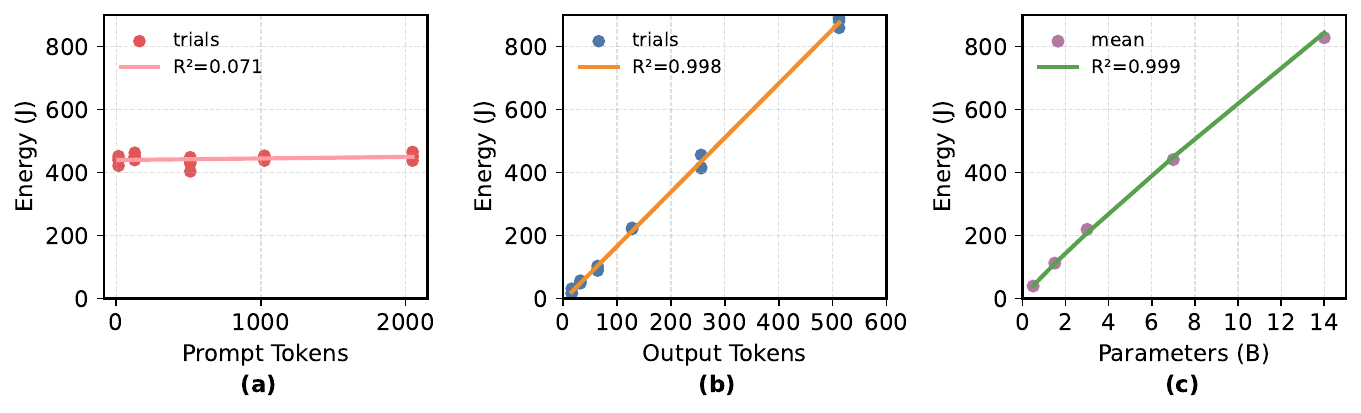}
\vspace{-28pt}
\caption{Energy characterization: (a) Energy vs. input length with fixed output (256 tokens), showing Prefill's minimal contribution; (b) Energy vs. output length with fixed input (64 tokens), demonstrating strong linear dependence;  (c) Energy scaling with model parameters under fixed workload (64-input, 256-output tokens).}
\label{fig:macro_energy}
\vspace{-13pt}
\end{figure}

\subsubsection{Energy vs. Workload}
\label{sec:energy_workload}

To dissect the energy contributions of Prefill and Decode phases, we systematically varied input and output lengths while measuring total energy consumption. 

When fixing output length at 256 tokens and increasing input length from 16 to 2048 tokens (Fig.~\ref{fig:macro_energy}(a)), total net energy remains nearly constant at $\approx 440$ J (fluctuation within $\pm 2\%$), exhibiting extremely low correlation with input length ($R^2=0.071$). This leads to our first energy insight:
\insight{\label{ins:energy_decode_dominant}%
LLM inference energy consumption is dominated by the Decode phase, with Prefill contributing negligibly to total energy.} Although Prefill is computationally intensive, its one-time execution contributes minimal energy compared to the hundreds of memory-access-dominated iterations in Decode phase, where KV Cache operations accumulate to dominate total energy. 

Conversely, when fixing input length and varying output length (Fig.~\ref{fig:macro_energy}(b)), we observe a fundamentally different pattern:
\insight{\label{ins:energy_linear_output}%
Energy consumption exhibits a highly linear relationship with the number of output tokens ($R^2=0.999$).} The fitted relationship $E = 1.72 \times L_{out} - 6.48$ reveals a marginal energy cost of approximately 1.72 J per generated token. The small intercept ($\approx6.5$ J) further confirms the minimal energy contribution of Prefill. 

%These results establish that LLM inference energy is primarily determined by Decode and can be accurately modeled as a linear function of output token count.

\subsubsection{Energy vs. Model Scale}
\label{sec:energy_scale}

We further investigate the relationship between energy consumption and model scale by measuring total net energy across models ranging from Qwen2.5-0.5B to 14B under a fixed workload (64 input tokens, 256 output tokens).

As shown in Fig.~\ref{fig:macro_energy}(c), energy consumption exhibits a strong dependence on model scale:
\insight{\label{ins:energy_model_scale}%
Energy consumption scales near-linearly with model parameter count $P$ ($R^2 = 0.998$).}
The experimental data is well-characterized by both a power-law fit ($E = 76.6 \times P^{0.91}$) with exponent close to 1, and a linear relationship ($E = 28.2 + 57.7P$). The observed near-linear energy scaling stems from larger models concurrently elevating instantaneous power draw and prolonging execution time, with their product dictating the total energy consumption.

\subsubsection{Predictability of Inference Energy}
\label{sec:energy_predictability}
Our analysis establishes that \insight{\label{ins:energy_prompt_predictable}%
LLM inference energy consumption can be accurately predicted from prompt features prior to execution.}
This capability derives from a composite function: prior work~\cite{zheng2023response} enables output length prediction from prompts ($L_{out} \approx g(\text{Prompt})$), while our results confirm a strong linear energy-length relationship ($E \approx f(L_{out})$). Thus, $E \approx f(g(\text{Prompt}))$ provides early energy estimation for incoming requests.

\section{Root Cause Explanation: Determining the "Why" – Microarchitectural Etiology} \label{sec:micro_level_characterization}

%Having established the macroscopic two-phase heterogeneity, we now investigate its underlying hardware origins. Through Roofline modeling, we quantify the fundamental divergence in arithmetic intensity between Prefill (high-intensity, compute-bound) and Decode (low-intensity, memory-bound) phases. Warp stall analysis further reveals distinct bottleneck patterns: Prefill execution is dominated by pipeline busy and execution dependency stalls, indicating computational saturation, while Decode shows predominant memory dependency stalls, reflecting KV-cache access bottlenecks. Cache behavior examination completes this picture, showing significantly degraded L2 hit rates in Decode due to limited data reuse compared to Prefill's compute-friendly access patterns. Together, these microarchitectural insights provide a causal explanation for the observed step dichotomy, connecting macroscopic performance phenomena to their hardware-level determinants.

While above analysis established the pronounced two-phase heterogeneity in LLM inference, these findings remain at the level of aggregate performance metrics. Fundamental questions persist: which underlying operators drive these phase differences, and how are GPU execution resources utilized—or wasted—in each phase? To uncover the microarchitectural mechanisms behind the macroscopic observations, this section systematically investigates the computational roots of phase divergence. We first identify and quantify the core CUDA kernels dominating execution time, then employ Roofline analysis to diagnose their resource constraints, followed by warp stall analysis to pinpoint hardware-level inefficiencies, and finally examine memory access patterns.

\begin{figure}[!b]
\vspace{-15pt}
\centering
\includegraphics[width=1\textwidth]{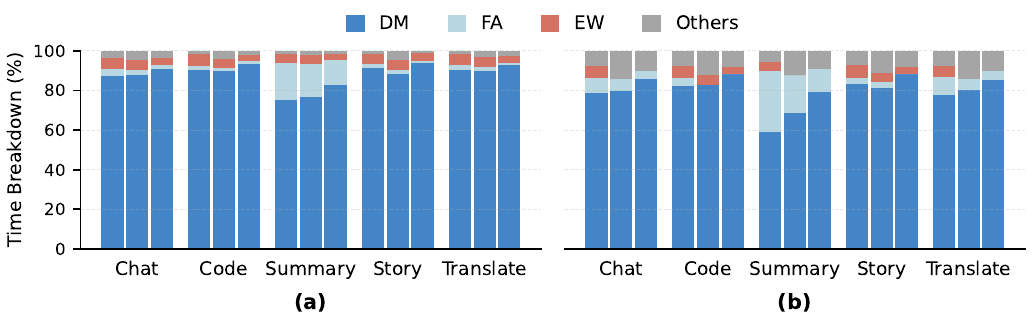}
\vspace{-25pt}
\caption{Consistent dominance of Dense Matrix (DM) kernels in both (a) Prefill and (b) Decode phases, revealing similar computational structures despite divergent performance characteristics. In each scenario group, the three bars represent Llama-3-8B, Qwen2.5-7B, and Qwen2.5-32B respectively from left to right.}
\label{fig:micro_level_kernel_distribution}
%\vspace{-15pt}

\end{figure}

\subsection{Execution Bounds Analysis}
\label{sec:execution_bounds_analysis}

This subsection identifies and categorizes the CUDA kernels responsible for the majority of execution time, analyzing their computational characteristics and hardware resource constraints.

\subsubsection{Kernel Classification and Time Distribution}
We focus on the top-5 most time-consuming CUDA kernels, which collectively account for >90\% of Prefill and >85\% of Decode execution time. Analyzing these critical kernels reveals the core computational patterns in each phase. 

%As shown in Table~\ref{tab:kernel_category}, 
We classify these dominant kernels into three categories based on their computational roles in LLM architectures:
\insight{\label{ins:kernel_classification}%
The top-5 kernels fall into three categories: DM (Dense Matrix), FA (Fused Attention), and EW (Element-wise).} 
The DM comprises dense matrix multiplications (GEMM) for FFN (gate-up, down) and attention (Q/K/V, output) projections. The FA includes fused attention kernels (e.g., FlashAttention) that minimize memory movement. The EW covers element-wise operations like activation, normalization, and positional encoding.

%The DM category executes dense matrix multiplications that form the computational backbone of Transformer architectures, including FFN projections (gate-up, down) and attention projections (Q/K/V, output). The FA category encompasses fused attention kernels like FlashAttention that combine multiple attention steps to minimize memory movement and kernel launch overhead. The EW category handles element-wise operations with typically low arithmetic intensity, including activations, normalization, and positional encoding.

From Fig.~\ref{fig:micro_level_kernel_distribution}(a) and (b), we observe a striking pattern:
\insight{\label{ins:kernel_distribution_similarity}%
Despite fundamental performance differences, Prefill and Decode share remarkably similar kernel-type distributions, with DM kernels dominating execution time in both phases.} This distribution similarity despite phase heterogeneity indicates that the core computational patterns remain consistent, while kernel efficiency and resource utilization differ substantially.

\begin{figure}[!t]
\vspace{-10pt}

\centering
\includegraphics[width=0.9\textwidth]{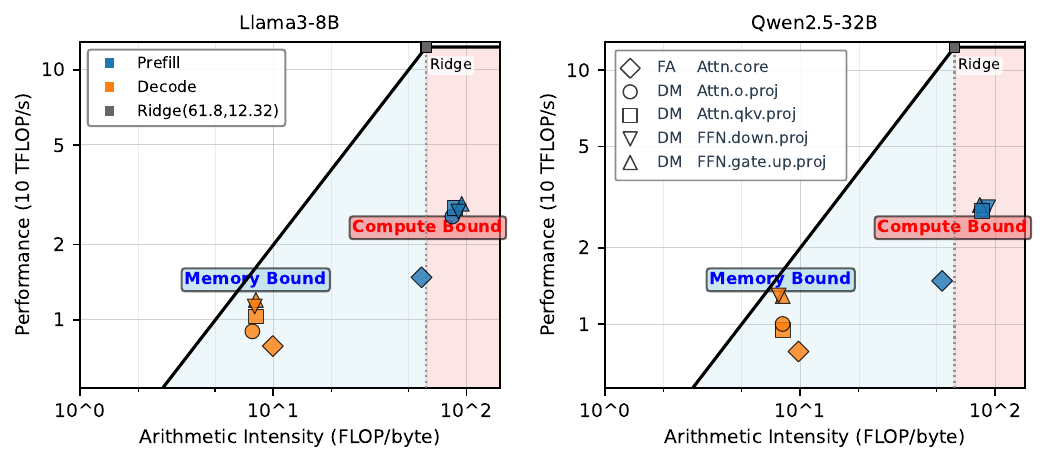}
\vspace{-15pt}
\caption{Roofline analysis of major kernels under the Chat scenario, measured on Tensor Cores.}\label{fig:roofline_comparison_llama3-8b_vs_qwen2.5-32b}
\vspace{-12pt}
\end{figure}

\subsubsection{Roofline Analysis.}

The Roofline model relates arithmetic intensity (AI) to achievable performance, offering an intuitive method to diagnose whether an operator is compute-bound or memory-bound. Fig.~\ref{fig:roofline_comparison_llama3-8b_vs_qwen2.5-32b}(a) and (b) illustrates the Roofline distributions of key kernels for two representative models—Llama3-8B and Qwen2.5-32B—under the Chat scenario.

%Across models of varying scales, Prefill and Decode phases exhibit systematic separation in Roofline space. Prefill kernels (blue markers) generally achieve high arithmetic intensity (AI $\approx$ 55$\sim$100), residing near or within the compute-bound region, which reflects dense GEMM-dominated execution with high Tensor Core utilization. In contrast, Decode kernels (orange markers) cluster in the low-AI regime (AI $\approx$ 1$\sim$10), where performance is constrained by memory bandwidth—a consequence of frequent KV-cache accesses and reduced computational density. These Roofline results show: \insight{\label{ins:compute_vs_memory_bound} Prefill is fundamentally compute-bound, while Decode is memory-bound.}

Prefill and Decode phases are systematically separated in Roofline space. Prefill kernels (high AI, $\approx$55–100) are compute-bound, reflecting dense GEMM execution, whereas Decode kernels (low AI, $\approx$1–10) are memory-bound due to frequent KV-cache accesses. Thus: \insight{\label{ins:compute_vs_memory_bound} Prefill is fundamentally compute-bound, while Decode is memory-bound.}
A detailed breakdown by kernel category further elucidates this behavior. \uline{DM kernels} in Prefill (e.g., \texttt{FFN.gate.up.proj}, \texttt{Attn.qkv.proj}) achieve high AI and performance. During Decode, however, their AI drops sharply due to small per-token inputs, shifting execution toward memory-bound regimes. \uline{FA kernels} such as \texttt{Attn.core} exhibit moderate AI in Prefill, balancing compute and memory demands. 
In Decode, repeated KV-cache access further reduces AI, placing them firmly in the memory-bound region. Notably, under long-context workloads such as the Summary scenario, the quadratic complexity of Attention significantly increases AI—reaching 319.3 for Llama3-8B and 382.1 for Qwen2.5-32B in the Prefill phase—thereby pushing Attention kernels deeper into the compute-bound zone. \insight{\label{ins:roofline_phase_heterogeneity}%
Phase heterogeneity originates from tensor size and data reuse differences: Prefill leverages large tensors with high reuse, while Decode operates on small per-token inputs with frequent memory access, determining their compute-bound vs. memory-bound characteristics.}

\begin{figure}[!t]
\vspace{-10pt}

    \centering
    \includegraphics[width=1\textwidth]{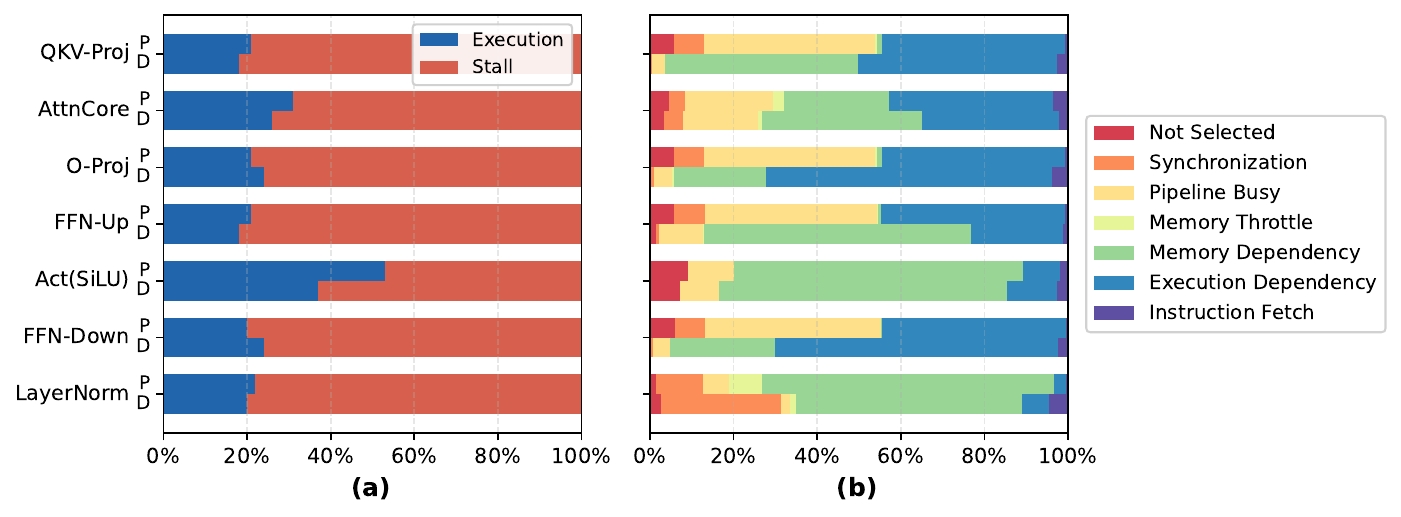}
    \vspace{-25pt}
    \caption{
        Issue stall analysis for \textbf{Llama3-8B} inference in the \textbf{Chat} scenario.
(a) Proportion of stall time to total execution time;
(b) Detailed breakdown of stall causes across Prefill and Decode phases.
         \footnotesize{
            Abbreviations: 
            QKV-Proj: Query/Key/Value projection; 
            AttnCore: Core attention computation (e.g., FlashAttention); 
            O-Proj: Output projection of Attention;
            FFN-Up: Up-projection and gating fusion in the feed-forward network; 
            Act (SiLU): Activation and element-wise multiplication in FFN;
            FFN-Down: Down-projection layer in FFN; 
            LayerNorm: Layer normalization;
            P: Prefill; D: Decode.
        }
    }
    \label{fig:kernel_stall_analysis}
        \vspace{-15pt}

\end{figure}

\subsection{Issue Stall Analysis}
\label{sec:issue_stall_analysis}
While Roofline analysis identifies resource bounds, GPU stall behavior explains the performance divergence between Prefill and Decode. Our warp-level analysis in Fig.~\ref{fig:kernel_stall_analysis}(a) reveals that both phases exhibit low GPU \textit{Execution} (Prefill: 27\%, Decode: 24\%), meaning 70\%--80\% of cycles are spent stalled. However, the root causes differ fundamentally between phases, as detailed in Fig.~\ref{fig:kernel_stall_analysis}(b).

%While Roofline analysis identifies the resource-bound characteristics of each phase, understanding GPU stall behavior and occupancy dynamics is essential to fully explain the performance divergence between Prefill and Decode. This subsection systematically examines instruction issue stalls during inference, analyzing their distribution and root causes across phases. Through warp-level stall analysis, we establish direct connections between the observed compute-bound and memory-bound behaviors and their underlying hardware manifestations.

%As shown in Fig.~\ref{fig:kernel_stall_analysis}(a), a striking observation emerges: GPU \textit{Execution\%} remains remarkably low in both phases (approximately 27\% in Prefill and 24\% in Decode), indicating that stalls consume 70\%--80\% of total execution time. This reveals that GPU hardware spends the majority of LLM inference cycles waiting rather than computing. However, the underlying causes of these high stall rates differ fundamentally between phases.

\subsubsection{Prefill Phase Stalls}

In Prefill, GEMM-intensive kernels including QKV-Proj, O-Proj, FFN-Up, and FFN-Down are predominantly stalled by Execution Dependency ($\approx$39\%) and Pipeline Busy ($\approx$37\%), as shown in Fig.~\ref{fig:kernel_stall_analysis}(b). This pattern uncovers the microarchitectural basis for Prefill's compute-bound nature:
\insight{\label{ins:prefill_dual_saturation}%
Prefill experiences dual saturation of computational throughput and instruction latency: high Pipeline Busy indicates sustained Tensor Core utilization, while Execution Dependency reflects warps waiting for long-latency HMMA instructions.}

The low Memory Dependency (24\%) confirms effective memory latency masking through high AI and data reuse. Attention kernels show balanced stall distributions, aligning with their memory-aware design, while element-wise operations like Act (SiLU) remain memory-dependent.

%The relatively low Memory Dependency (24\%) confirms that memory latency is effectively masked by high AI and data reuse. Attention kernels (e.g., FlashAttention) display more balanced stall distributions, consistent with their memory-aware design, while element-wise operations like \texttt{Act (SiLU)} exhibit memory-dependent behavior but contribute minimally to total runtime.

%Decode exhibits a fundamentally different stall profile. Pipeline Busy drops dramatically (to just 3$\sim$4\% for O-Proj) while Memory Dependency becomes the dominant bottleneck: \insight{\label{ins:decode_memory_bound}%
%Decode is primarily memory-dependent: limited parallelism and frequent KV-cache access make memory latency the overwhelming stall source.}

\subsubsection{Decode Phase Stalls}
Decode's stall profile shifts fundamentally: Pipeline Busy plunges to 3$\sim$4\%, while Memory Dependency dominates. This reveals that \insight{\label{ins:decode_memory_bound}%
Decode is primarily memory-bound, as limited parallelism and intensive KV-cache access make memory latency the primary bottleneck.}
This shift is particularly evident in Attention kernels (AttnCore), where the dominant stall transitions from Execution Dependency (32.8\% in Prefill) to Memory Dependency (32.7\% in Decode), consistent with intensive KV-cache operations. Further analysis reveals heterogeneous bottlenecks among Decode's GEMM kernels: FFN-Up is memory-bound, with Memory Dependency reaching 58\% due to GEMV-like computation patterns that cannot hide HBM weight-loading latency. O-Proj/FFN-Down are execution-bound, with Execution Dependency reaching 51\% due to extreme register pressure ($\approx$130 live registers) that limits schedulable warps and prevents HMMA latency hiding.
This reveals: \insight{\label{ins:decode_gemm_heterogeneous} Decode's GEMM kernels exhibit heterogeneous bottlenecks: FFN-Up is memory limited, while O-Proj/FFN-Down are instruction-dependency bound.}
Additionally, LayerNorm shows significantly increased Synchronization stalls (from 11.0\% to 27.7\%), revealing:
\insight{\label{ins:decode_sync_stall_smallbatch}%
Small-batch reduction operations in Decode exacerbate synchronization overhead, as early-finishing warps wait at barriers, creating tail latency effects.}

%This comprehensive stall analysis completes the microarchitectural explanation for phase heterogeneity, connecting high-level observations to specific hardware bottleneck mechanisms.

\subsection{Memory Pattern Analysis}
\label{sec:memory_pattern_analysis} 
\begin{figure}[!t]
    \vspace{-10pt}
    \centering
    \includegraphics[width=1.0\textwidth]{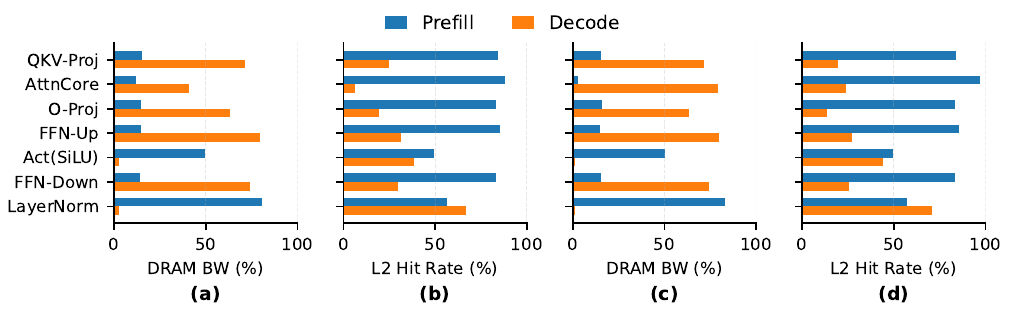}
    \vspace{-25pt}
    \caption{Memory performance for key kernels in Llama-3-8B. (a) DRAM Bandwidth and (b) L2 Hit Rate in Chat scenario; (c) DRAM Bandwidth and (d) L2 Hit Rate in Summary scenario.}
    \label{fig:memory_performance_combined}
    \vspace{-15pt}
\end{figure}

We complete our microarchitectural characterization by examining the memory behavior that distinguishes Prefill and Decode phases. Using DRAM bandwidth utilization and L2 cache hit rate as key metrics, we analyze kernel-level memory patterns across Chat (Fig.~\ref{fig:memory_performance_combined}(a) and (b)) and Summary (Fig.~\ref{fig:memory_performance_combined}(c) and (d)) scenarios in Llama-3-8B. Phase transitions are quantified via median per-kernel differences between Decode and Prefill, providing a robust measure of trend shifts.

%This subsection examines memory subsystem behavior—including DRAM access patterns and cache efficiency—to elucidate the memory characteristics distinguishing Prefill and Decode phases during LLM inference. By analyzing bandwidth utilization and cache performance across different scenarios, we complete the microarchitectural characterization of phase heterogeneity. 

%As shown in Fig.~\ref{fig:memory_performance_combined}, we characterize the memory behavior of key kernels in Llama-3-8B across Prefill and Decode phases through two critical metrics: DRAM bandwidth utilization and L2 cache hit rate. Panels (a) and (c) present results for the Chat scenario, while panels (b) and (d) correspond to the Summary scenario. To systematically quantify phase transitions, we compute per-kernel differences between Decode and Prefill, using the median value across kernels as a robust indicator of overall trends while mitigating the influence of outliers such as LayerNorm.

\subsubsection{Phase-Level Memory Characteristics} %Quantitative analysis reveals systematic differences in memory subsystem behavior between Prefill and Decode phases. These phase-level patterns establish the foundational memory characteristics that distinguish the two execution phases and provide context for subsequent operator-level analysis.

In the Chat scenario, median DRAM utilization increases by 48.2\% from Prefill to Decode, while median L2 hit rate decreases by 54.0\%. The Summary scenario shows even more pronounced shifts, with 56.2\% higher DRAM utilization and 58.5\% lower L2 hit rates. These consistent patterns lead to a key insight:
\insight{\label{ins:decode_dram_locality_shift}%
Compared to Prefill, Decode exhibits a clear shift toward higher DRAM bandwidth dependence and lower data locality.}
This fundamental divergence stems from intrinsic computational differences: Prefill leverages large GEMM operations with extensive data reuse, while Decode's token-by-token execution and frequent KV-cache accesses result in insufficient arithmetic intensity to hide memory latency.

\subsubsection{Operator-Level Memory Patterns} We analyze how memory behavior shifts across operators from Prefill to Decode.
Attention Kernels suffer severe locality degradation, with L2 hit rates dropping 81.9\% (Chat) and 73.3\% (Summary), while DRAM utilization increases up to 76.4\%. \insight{\label{ins:attncore_locality_degradation} Attention kernels transition from compute-friendly to memory-bound due to exhaustive KV-cache scanning.} Prefill's consolidated access maximizes reuse, whereas Decode's scattered reads across the KV-cache destroy locality, making performance bandwidth-bound.
FFN Kernels exhibit a bottleneck shift, with DRAM utilization rising 62\% and arithmetic intensity plummeting from $\sim$95 to $\sim$8 FLOP/byte. \insight{\label{ins:decode_ffn_weight_bandwidth} FFN transitions from compute-bound to memory-bound, with weight-loading bandwidth as the primary constraint.} Prefill's batch GEMM reuses weights, while Decode's per-token GEMVs reload weights from DRAM.
LayerNorm shows improved locality (DRAM utilization drops 78\%–82\%, L2 hit rates slightly increase). Its small working set fits in on-chip caches, preventing it from becoming a bottleneck.

\subsubsection{Scenario-Dependent Memory Behavior} We dissect how memory behavior varies across scenarios, focusing on how context length interacts with phase-specific memory characteristics.
Comparative analysis reveals contrasting trends between Chat and Summary scenarios: AttnCore DRAM utilization increases by an additional 38.1\% in Decode when transitioning to the long-context Summary scenario, while Prefill utilization decreases by 9.3\%. This divergence confirms:
\insight{\label{ins:long_context_attention_bottleneck}%
Extended context lengths systematically amplify the memory bottleneck in attention kernels.}
This occurs because Prefill leverages enhanced data reuse, while Decode's linearly growing KV-cache footprint intensifies bandwidth pressure, further entrenching its memory-bound nature.

%The underlying cause lies in phase-specific characteristics: Prefill benefits from enhanced data reuse in longer contexts, while Decode suffers from linearly growing KV-cache footprints that intensify DRAM pressure. Thus, longer contexts reinforce Decode's memory-bound nature while preserving Prefill's compute-bound profile.

%We examine how memory behavior varies across application scenarios, focusing particularly on how context length interacts with phase-specific memory characteristics. This analysis reveals how workload properties and architectural constraints jointly determine system performance.

%Comparative analysis between Chat and Summary scenarios reveals contrasting trends: AttnCore DRAM utilization increases by an additional 38.1\% in the Decode phase when transitioning from Chat to Summary, while Prefill shows a 9.3\% reduction in DRAM utilization. This divergence leads to a key insight:
%\insight{\label{ins:long_context_attention_bottleneck} Extended context lengths systematically amplify the memory bottleneck in attention kernels.} This phenomenon stems from fundamental phase characteristics: Prefill benefits from enhanced data reuse in longer contexts, whereas Decode suffers from linearly growing KV-cache footprints that dramatically intensify DRAM bandwidth pressure. Consequently, escalating context length reinforces Decode's memory-bound nature while preserving Prefill's compute-bound characteristics.

\section{Scaling Analysis: Validating the "How" – System-Level Governing Principles} \label{sec:system_level_characterization}

With the root causes established, we examine how these phase-aware principles govern performance in deployment scenarios. In multi-GPU environments, we demonstrate that the compute-intensive Prefill achieves optimal scaling through Tensor Parallelism (TP)~\cite{shoeybi2019megatron}, while the memory-intensive Decode benefits more from Pipeline Parallelism (PP)~\cite{huang2019gpipe} due to reduced communication overhead. When extending to edge devices, we observe that resource constraints amplify these phase-specific characteristics: memory bandwidth limitations exacerbate Decode bottlenecks, while thermal constraints trigger frequency throttling that particularly impacts Prefill performance. These scaling experiments validate that the two-phase paradigm persists across system scales.

%With the root causes established, we now examine how these phase-aware principles govern performance in realistic deployment scenarios. In multi-GPU environments, we demonstrate that the compute-intensive Prefill achieves optimal scaling through Tensor Parallelism~\cite{shoeybi2019megatron}, while the memory-intensive Decode benefits more from Pipeline Parallelism~\cite{huang2019gpipe} due to reduced communication overhead. When extending to edge devices, we observe that resource constraints systematically amplify these phase-specific characteristics: memory bandwidth limitations exacerbate Decode bottlenecks, while thermal constraints trigger frequency throttling that particularly impacts Prefill performance. These scaling experiments validate that the two-phase paradigm persists across system scales.

%and provides concrete guidelines for phase-aware system configuration in diverse deployment environments.

\subsection{Multi-GPU Scaling Laws}
\label{sec:parallel_strategy_analysis}

Building on the phase-aware scaling foundation, we evaluate parallelization strategies against the distinct computational profiles of Prefill and Decode in multi-GPU environments. We systematically compare tensor, pipeline, and hybrid parallelism. Using Qwen2.5-32B on four A100s, we test five configurations: Single (1-GPU baseline), TP2 (2-GPU tensor parallelism), PP2 (2-GPU pipeline parallelism), TP4 (4-GPU tensor parallelism), PP4 (4-GPU pipeline parallelism), and TP2+PP2 (4-GPU hybrid parallelism with two TP2 groups forming pipeline stages). Two workloads represent key scenarios: Chat (64-input, 128-output) for Decode-dominant and Summary (4096-input, 16-output) for Prefill-dominant patterns, enabling a comprehensive study of phase-dependent parallelization.

%Having established the theoretical foundation for phase-aware scaling, we now empirically evaluate how different parallelization strategies interact with the distinct computational characteristics of Prefill and Decode phases. This analysis focuses on multi-GPU environments, where the trade-offs between computational parallelism and communication overhead become particularly pronounced. We systematically compare tensor parallelism, pipeline parallelism, and hybrid approaches across representative workloads to derive practical guidelines for phase-aware system configuration.

%We evaluate five parallelization configurations on a four-A100 platform using Qwen2.5-32B: Single (1-GPU baseline), TP2 (2-GPU tensor parallelism), PP2 (2-GPU pipeline parallelism), TP4 (4-GPU tensor parallelism), PP4 (4-GPU pipeline parallelism), and TP2+PP2 (4-GPU hybrid parallelism with two TP2 groups forming pipeline stages). Our evaluation employs two representative workloads: Chat (64-token input, 128-token output) representing Decode-dominant scenarios, and Summary (4096-token input, 16-token output) representing Prefill-dominant scenarios, enabling comprehensive examination of phase-dependent parallelization behavior.

\begin{table}[t]
\vspace{-10pt}

\centering
\caption{Performance comparison of different parallelization strategies across two representative workloads: Chat with short input and long output; Summary with long input and short output.}
\label{tab:multi_gpu_comparison}
\vspace{-10pt}
\resizebox{0.8\textwidth}{!}{
\begin{tabular}{|c|c|c|c|c|c|c|c|}
\hline
Workload & {Metric} & {Single} & {TP2} & {PP2} & {TP4} & {PP4} & {TP2+PP2} \\ \hline
\multirow{2}{*}{Chat}   & {Throughput (tokens/s)} & 22.16 & 20.53 & 21.08 & 18.55 & 18.54 & 18.18 \\  
& {Latency (ms/token)} & 45.12 & 48.72 & 47.43 & 53.92 & 53.94 & 55.01 \\
\hline
\multirow{2}{*}{Summary}  & { Throughput (tokens/s)} & 8.19 & 11.21 & 8.22 & 8.77 & 7.40 & 10.65 \\
& { Latency (ms/token)} & 122.16 & 89.19 & 121.67 & 114.06 & 135.17 & 93.94 \\
\hline
\end{tabular}
}
\vspace{-10pt}

\end{table}

\begin{comment}

\begin{figure}[!t]
\centering
\includegraphics[width=0.95\textwidth]{figures/system-level/multi_gpu_comparison.pdf}
\vspace{-15pt}
\caption{Performance comparison of different parallelization strategies across two representative workloads: (a) Chat with short input and long output; (b) Summary with long input and short output. }
\label{multi_gpu_comparison}
\end{figure}

\begin{table}[h]
\centering
\caption{Performance comparison of parallelization strategies across workloads.}
\label{tab:multi_gpu_comparison}
\begin{tabular}{lccc}
\toprule
\textbf{Workload} & \textbf{Strategy} & \textbf{Throughput (tokens/s)} & \textbf{Latency (ms/token)} \\
\midrule
\multirow{6}{*}{Chat} & Single & 22.16 & 45.12 \\
& TP2 & 20.53 & 48.72 \\
& PP2 & 21.08 & 47.43 \\
& TP4 & 18.55 & 53.92 \\
& PP4 & 18.54 & 53.94 \\
& TP2+PP2 & 18.18 & 55.01 \\
\hline
\multirow{6}{*}{Summary} & Single & 8.19 & 122.16 \\
& TP2 & 11.21 & 89.19 \\
& PP2 & 8.22 & 121.67 \\
& TP4 & 8.77 & 114.06 \\
& PP4 & 7.40 & 135.17 \\
& TP2+PP2 & 10.65 & 93.94 \\
\bottomrule
\end{tabular}
\end{table}
\end{comment}

\begin{figure}[!t]
\centering
\includegraphics[width=0.90\textwidth]{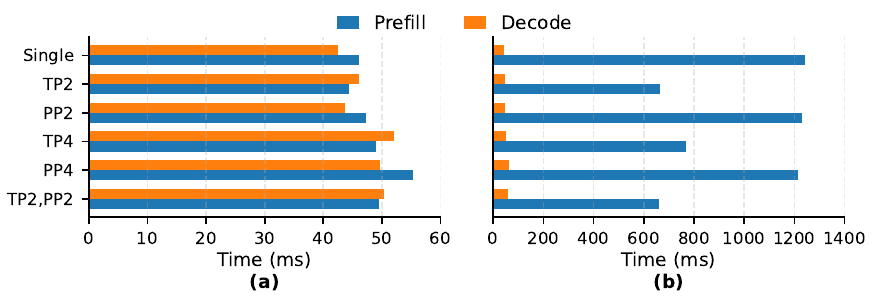}
\vspace{-15pt}
\caption{Prefill and decode (per-step) latency under different parallelization strategies across two representative workloads: (a) Chat with short input and long output; (b) Summary with long input and short output.}
\label{multi_gpu_time_comparison}
\vspace{-20pt}
\end{figure}

\subsubsection{Overall Performance}
\label{sec:overall_performance}

We evaluate end-to-end performance across workloads to establish how parallelization interacts with phase-specific patterns, with results shown in Table~\ref{tab:multi_gpu_comparison} and Fig.~\ref{multi_gpu_time_comparison}.

\uline{Chat Scenario: Decode-Dominant.} Single-GPU achieves optimal throughput (22.2 tokens/s), while parallel configurations reduce it (TP2: 20.5, PP2: 21.1) and increase latency (45.1 ms/token to 53.9–55.0 ms/token). Phase-level analysis in Fig.~\ref{multi_gpu_time_comparison}(a) reveals that while TP2 slightly reduces Prefill latency (46.1 ms → 44.4 ms), Decode latency consistently increases due to communication and synchronization overhead. This leads to a critical finding:
\insight{\label{ins:decode_single_gpu_better}%
For Decode-dominant workloads, communication and synchronization costs outweigh computational parallelism benefits, making single-GPU execution optimal for both latency and throughput.}

\uline{Summary Scenario: Prefill-Dominant.} Tensor parallelism excels: TP2 boosts throughput by 36.6\% (11.2 tokens/s) and cuts latency by 27\% (89.2 ms/token). TP dramatically shortens Prefill (1.24 s → 0.66 s) as shown in Fig.~\ref{multi_gpu_time_comparison}(b). Pipeline parallelism shows minimal gains (PP2: 8.2 tokens/s) or degradation (PP4: 7.4 tokens/s), establishing: \insight{\label{ins:prefill_tp_scaling} Prefill-dominant workloads benefit substantially from tensor parallelism, which effectively distributes GEMM computation, while pipeline parallelism suffers from stage serialization and bubble overhead.}

These results reveal fundamental phase-dependent scaling, leading to a crucial design insight: \insight{\label{ins:phase_aware_parallelism} Optimal parallelization is phase-aware: Prefill scales with tensor parallelism, while Decode performs best on single-GPU or tuned pipeline parallelism.}

\subsubsection{Time Breakdown}
\label{sec:execution_time_breakdown}
\begin{figure}[!t]
\vspace{-10pt}
\centering
\includegraphics[width=0.9\textwidth]{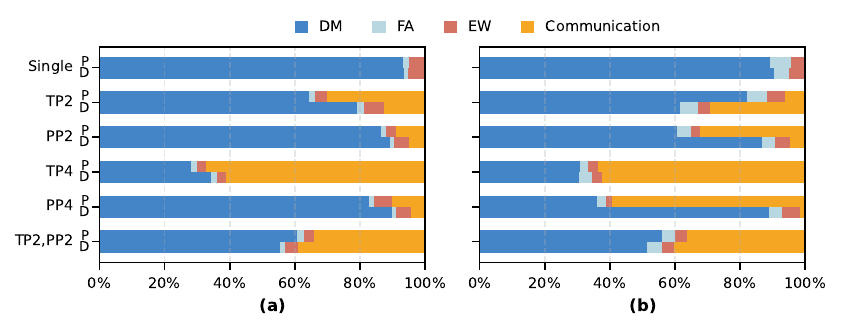}
\vspace{-18pt}
\caption{
Kernel-level time decomposition under different parallelization strategies:
(a) Chat scenario (Decode-dominated workload) and
(b) Summary scenario (Prefill-dominated workload).
P and D denote Prefill and Decode phases respectively.
Stacked bars show the proportional time spent in DM (Dense Matrix), FA (Fused Attention), EW (Element-Wise), and Communication operations.
}
\label{multi_gpu_kernel_analysis}
\vspace{-15pt}

\end{figure}

To elucidate performance variations, we decompose kernel time into DM, FA, EW, and Communication, with breakdowns for both Chat and Summary shown in Fig.~\ref{multi_gpu_kernel_analysis}.

\uline{Chat Scenario: Decode-Dominant.}
Single-GPU execution is dominated by DM kernels (>93\%), with FA and EW contributing 7\%. Multi-GPU configurations diverge sharply: TP introduces substantial Communication overhead (30\% in Prefill, >60\% in Decode for TP4), while PP maintains high DM share (83–90\%) with low Communication (5–10\%). The hybrid TP2+PP2 shows intermediate Communication levels. This divergence stems from communication patterns: TP requires frequent collective operations (e.g., All-Reduce) per layer, and in Decode-dominant workloads, minimal per-step computation cannot amortize this overhead, making communication dominant. PP, with only infrequent activation transfers at stage boundaries, preserves computational efficiency.

\uline{Summary Scenario: Prefill-Dominant.} TP2 introduces modest Communication (6\%) while drastically reducing Prefill latency (1.24 s → 0.66 s) by parallelizing DM kernels. PP2, however, incurs high Communication (32\%) without proportional computational benefit, maintaining similar Prefill latency (1.23 s). Aggressive configurations (TP4/PP4) become communication-dominant (59–64\%).

Our analysis establishes: \insight{\label{ins:kernel_comm_balance} Multi-GPU scaling efficiency is governed by the computation-communication balance.} Tensor parallelism justifies its cost in compute-intensive Prefill but fails in Decode, where per-step computation is too small to amortize overhead.

\subsection{Edge-Scaling Characteristics}
\label{sec:edge_device_profiling}

Edge devices present fundamentally different computational constraints compared to data-center GPUs, characterized by limited compute capability, reduced memory bandwidth, and strict power budgets. To understand how LLM inference behavior scales under these constraints, we deploy Llama3-8B on an NVIDIA Jetson AGX Orin platform. By contrasting Roofline distributions and SM frequency characteristics with our A100 findings, we characterize how performance bottlenecks transform when the same model executes on resource-constrained edge systems.

\begin{figure}[!t]
\vspace{-10pt}

\centering
\includegraphics[width=1\textwidth]{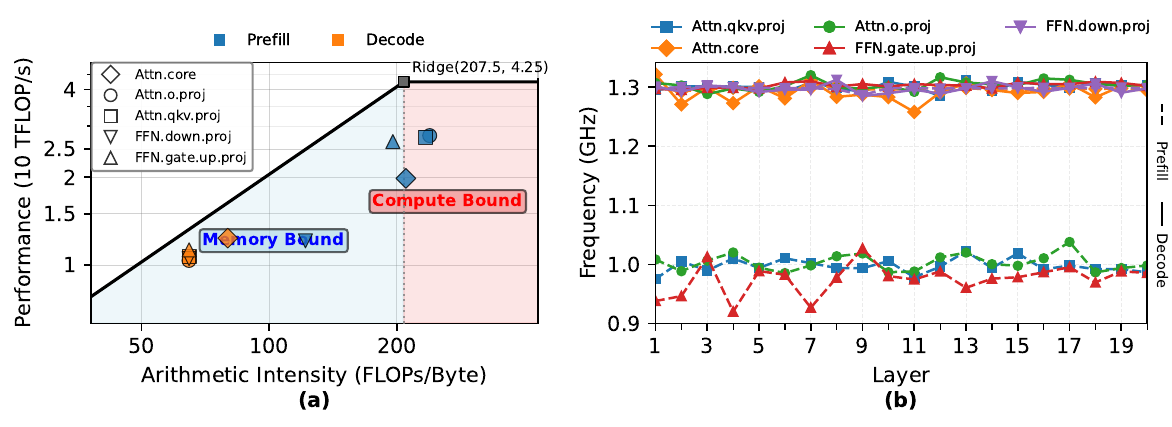}
\vspace{-25pt}
\caption{Computational characteristics of Llama3-8B on Jetson AGX Orin:
(a) Roofline analysis showing AI versus achieved performance for key kernels across Prefill and Decode;
(b) SM frequency distribution across transformer layers under 1.3 GHz frequency cap, revealing phase-dependent DVFS behavior.}
\label{jetson_roofline_freq}
\vspace{-15pt}
\end{figure}

\subsubsection{Kernel-level Roofline}
We examine how computational patterns translate to edge devices via Roofline analysis (Fig.~\ref{jetson_roofline_freq}), revealing how constraints reshape performance boundaries.

\uline{Prefill Phase.} Self-attention kernels (QKV-Proj, O-Proj, FlashAttention) maintain high AI near the compute-bound region, indicating stable computational characteristics across platforms. In contrast, FFN GEMM kernels shift dramatically: limited cache and LPDDR bandwidth force frequent data movement, reducing AI and causing a fundamental bottleneck transformation: \insight{\label{ins:edge_ffn_memory_bound} FFN kernels transition from compute-bound on data-center GPUs to memory-bound on edge devices, exhibiting heightened sensitivity to memory bandwidth and cache resources.}

\uline{Decode Phase.} Both attention and FFN kernels cluster strongly in the bandwidth-limited region with low AI. The combination of small per-step computation and intensive KV-cache access creates overwhelming memory pressure: \insight{\label{ins:jetson_kv_memory_limit} Decode performance becomes dominated by KV-cache data movement, exhibiting extreme memory-bandwidth sensitivity that amplifies the Prefill--Decode bottleneck divergence.}

Collectively, LLM inference in edge devices exhibits systematically stronger memory-bound characteristics compared to data-center environments, especially impacting FFN and Decode.

\subsubsection{Frequency Behavior and DVFS Effects} 
Building on Roofline analysis, we examine how dynamic voltage and frequency scaling (DVFS) interacts with phase-specific patterns. Fig.~\ref{jetson_roofline_freq}(b) presents SM frequency distributions of key kernels under Jetson's 1.3 GHz frequency cap.

%Building upon the Roofline analysis, we now examine how dynamic voltage and frequency scaling (DVFS) interacts with phase-specific computational patterns to shape performance and power characteristics on resource-constrained edge devices. Fig.~\ref{jetson_roofline_freq}(b) presents SM frequency distributions of key kernels across Prefill and Decode phases under Jetson's 1.3 GHz frequency cap. 

\uline{Prefill Phase.} In Prefill, compute-intensive GEMM kernels (QKV-Proj, O-Proj, Gate-Up-Proj) operate at significantly reduced average frequencies compared to the maximum cap, while FlashAttention maintains frequencies closer to the maximum due to its more balanced compute-memory profile. This frequency throttling indicates that on power-constrained edge devices, compute-bound kernels trigger aggressive DVFS mechanisms, sacrificing peak performance to manage instantaneous power consumption and thermal dissipation.

\uline{Decode Phase.} In contrast, Decode phase exhibits substantially different frequency behavior. Dominated by memory-intensive KV-cache operations rather than computational workloads, Decode maintains smoother frequency profiles that remain closer to the maximum cap across layers. The reduced computational intensity generates lower instantaneous power demand, minimizing thermal pressure and DVFS-induced frequency reductions.

This frequency analysis reveals a crucial dimension of phase heterogeneity:
\insight{\label{ins:edge_dvfs_heterogeneity}%
Edge devices exhibit phase-dependent DVFS behavior: compute-intensive Prefill triggers significant frequency throttling to manage power/thermal constraints, while memory-bound Decode maintains higher frequencies due to reduced computational pressure.} 
Prefill thus faces dual computational and power constraints, while Decode remains memory-subsystem limited.

%These findings have direct implications for edge system design. Prefill execution faces dual constraints of computational throughput and power/thermal limits, where excessive batch sizes or model dimensions can exacerbate DVFS-induced performance degradation. Decode, while computationally underutilized, remains fundamentally limited by memory subsystem performance. 

\section{Redefining Boundaries: Exploring the "Where" – Paradigm Shifts} \label{sec:design_aware_characterization}

Finally, we investigate how emerging model architectural and workflow innovations are redefining LLM inference performance boundaries. For model architecture, we characterize how MoE designs decouple total parameter count from computational cost via sparse activation, reducing latency while introducing new routing overheads that become pronounced in Decode. In workflow integration, we reveal how RAG shifts system bottlenecks from GPU computation to CPU-side memory access during retrieval operations, creating a new heterogeneous execution paradigm. These investigations demonstrate that while the two-phase perspective remains relevant, new architectures and workflows are actively reshaping where and how performance bottlenecks manifest.

\subsection{Tradeoffs in MoE Architecture}
\label{sec:moe_vs_dense}

We explore  MoE tradeoffs and how architecture transforms inference bottlenecks by comparing dense and MoE models across five scenarios. Measurements of Prefill/Decode latency and kernel-level time distributions reveal how sparse activation introduces distinct overhead patterns.

\begin{figure}[!t]
\vspace{-10pt}
\centering
\includegraphics[width=1\textwidth]{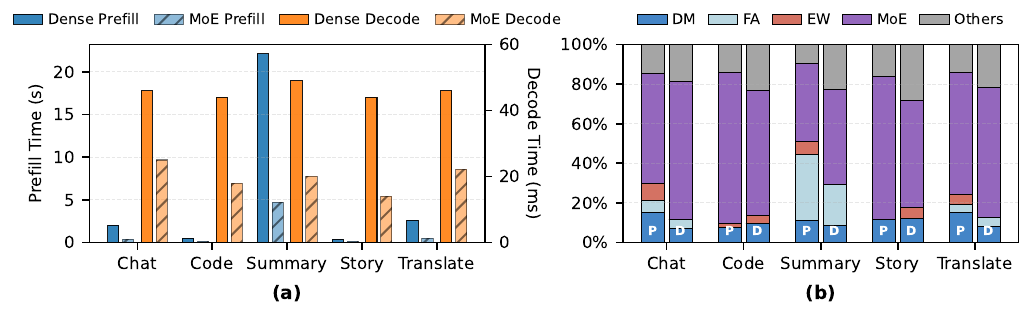}
\vspace{-25pt}
\caption{
Performance comparison and kernel-level analysis of dense versus MoE architectures:
(a) Prefill and per-step Decode latency comparison between Qwen2.5-32B (dense) and Qwen3-30B-A3B (MoE);
(b) Kernel time decomposition for the MoE model across five application scenarios.
}
\label{fig:moe_vs_dense_comparison}
\vspace{-15pt}
\end{figure}

\uline{Performance-Overhead Tradeoffs.} To uncover MoE's performance-cost trade-offs, we examine kernel-level time distributions. Our analysis compares end-to-end latency between architectural paradigms (Fig.~\ref{fig:moe_vs_dense_comparison}(a)) and decomposes MoE execution (Fig.~\ref{fig:moe_vs_dense_comparison}(b)) using DM/FA/EW categories augmented with ``MoE'' (expert operation) and ``Others'' (routing and auxiliary operations).
The model achieves significant speedups (4.56$\times$ Prefill, 2.39$\times$ Decode) over dense equivalents, revealing:  \insight{\label{ins:moe_tradeoff} MoE architectures decouple model capacity from computational cost, with inference performance determined by activated parameters rather than total parameter count.}
Time decomposition shows Prefill is dominated by expert operation (39.4\%--76.0\%) with minimal routing overhead (9.3\%--16.3\%). In Decode, expert operation remains substantial (48.0\%--69.8\%) but routing overhead surges to 18.6\%--28.1\%, becoming the second-largest cost. This phase-dependent pattern arises because Prefill's parallelization amortizes routing latency, while Decode's sequential generation exposes it, establishing: \insight{\label{ins:moe_decode_overhead_routing} MoE exhibits phase-dependent patterns: Prefill is dominated by expert operations, while Decode incurs substantial routing overhead that becomes a major performance factor alongside expert computation.}

\begin{figure}[!t]
%\vspace{-10pt}
\centering
\includegraphics[width=1\textwidth]{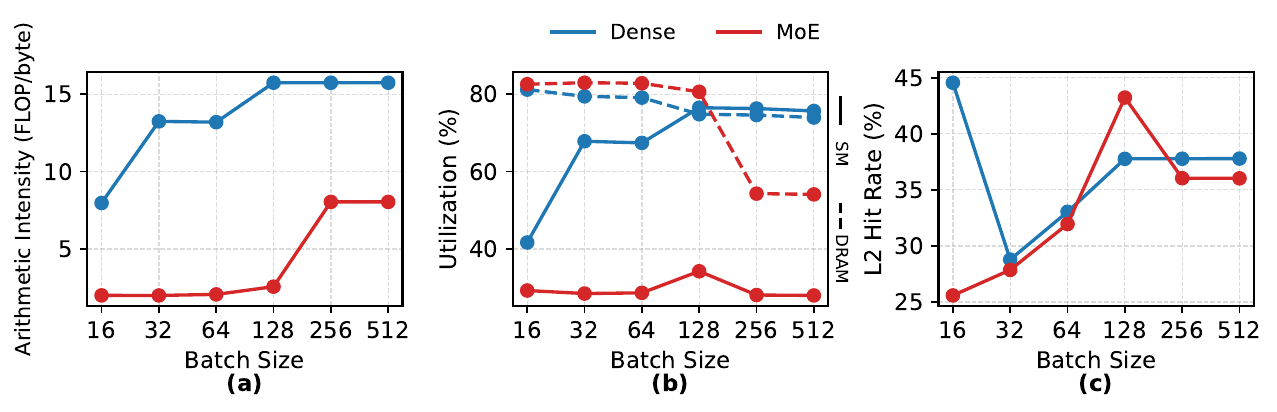}
\vspace{-25pt}
\caption{
Decode-phase FFN efficiency comparison under varying batch sizes:
(a) Arithmetic intensity evolution;
(b) SM and DRAM utilization patterns;
(c) L2 cache behavior for dense versus MoE architectures.
}
\label{fig:decode_ffn_phase_moe_vs_dense_triptych}
\vspace{-10pt}
\end{figure}

\uline{FFN Efficiency Tradeoffs.} We now focus on Decode-phase FFN efficiency, comparing dense and MoE models across batch sizes in: arithmetic intensity (Fig.~\ref{fig:decode_ffn_phase_moe_vs_dense_triptych}(a)), SM and DRAM utilization (Fig.~\ref{fig:decode_ffn_phase_moe_vs_dense_triptych}(b)), and cache behavior (Fig.~\ref{fig:decode_ffn_phase_moe_vs_dense_triptych}(c)). Arithmetic intensity reveals a fundamental gap: the dense model's FFN intensity grows to 15.74 FLOP/byte, while MoE remains low (8 FLOP/byte) due to \insight{\label{ins:moe_small_effective_batch} MoE routing distributes global batches across experts, creating small per-expert effective batches that prevent computational intensity from reaching dense model levels, fundamentally constraining GEMM efficiency.} 
Hardware utilization further illuminates the divergence: dense FFNs scale strongly toward compute-bound operation (SM utilization 41.7\%→76.5\%, decreasing DRAM pressure), while MoE FFNs sustain both low SM utilization (28\%–34\%) and high DRAM pressure (>80\% at small batches). This reveals: \insight{\label{ins:moe_ffn_memory_constraints} MoE FFNs operate under simultaneous computational underutilization and memory bandwidth saturation, demonstrating how sparse expert partitioning trades computational efficiency for model capacity.}
Cache analysis completes the picture and reveals MoE's batch-sensitive locality: at small batches, L2 hit rates are poor (25.6\% vs. dense 44.6\%) due to fragmented expert weight reuse; however, at larger batches (128), it surpasses dense models (43.2\% vs. 37.8\%) as token concentration within experts improves. This demonstrates: \insight{\label{ins:moe_l2_locality_batch_sensitive} MoE cache efficiency exhibits strong batch size dependence, requiring substantial concurrency to achieve effective weight reuse through expert-focused token routing.}

Collectively, MoE Decode-phase FFNs face compounded challenges from fragmentation, bandwidth saturation, and sensitive cache behavior.

\subsection{Workflow Design: RAG-in-the-Loop}
\label{sec:rag_workflow}

Retrieval-augmented generation~\cite{lewis2020retrieval} has become a standard paradigm to incorporate external knowledge. While prior analysis focused on GPU-centric inference, we examine how integrating retrieval reshapes system bottlenecks. We implement a hybrid RAG workflow using LightRAG~\cite{guo-etal-2025-lightrag}.

\begin{figure}[!t]
% \vspace{-10pt}
\centering
\includegraphics[width=1\textwidth]{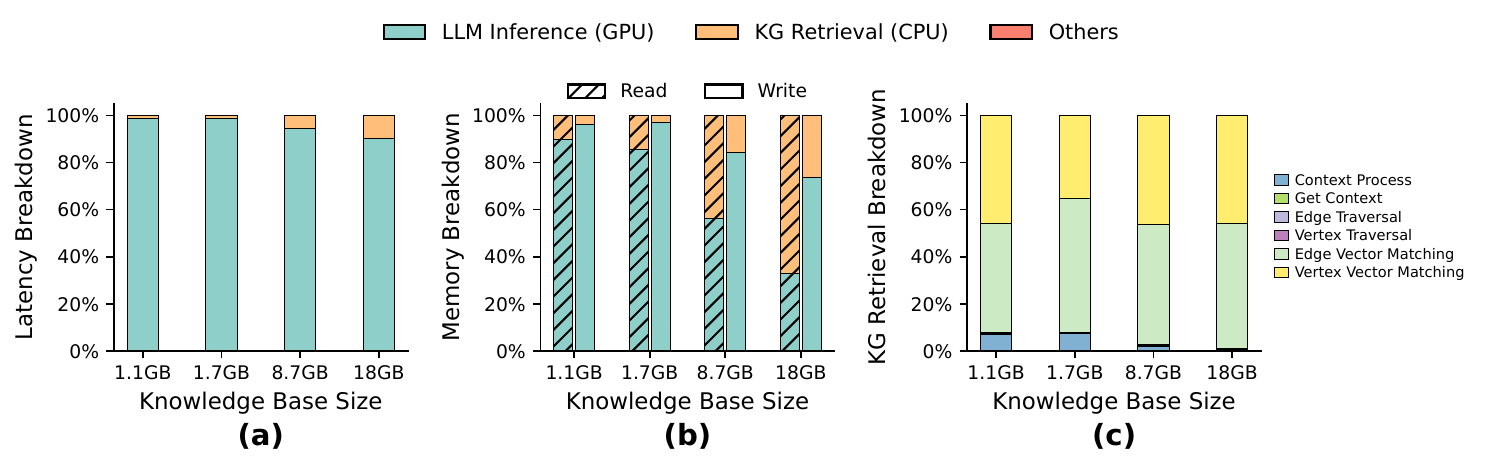}
\vspace{-25pt}
\caption{Characterizing RAG scalability: (a) Latency distribution between GPU inference and CPU retrieval across knowledge base sizes; (b) Memory traffic breakdown; (c) Detailed time breakdown of KG retrieval.}
\label{fig:rag_macro_breakdown}
\vspace{-15pt}
\end{figure}

\uline{End-to-End Workflow Scalability.} Fig.~\ref{fig:rag_macro_breakdown}(a) shows workload evolution with knowledge base scale: CPU-side KG retrieval incurs increasingly higher latency share as the knowledge base grows (1.1 GB to 18 GB), becoming non-negligible at large scales. Memory analysis (Fig.~\ref{fig:rag_macro_breakdown}(b)) confirms this shift, showing KG retrieval dramatically increases CPU-side memory traffic while GPU patterns remain stable. This reveals: \insight{\label{ins:rag_kg_memory_constraint} Scaling knowledge bases triggers a bottleneck shift from GPU computation to CPU-side retrieval and memory overhead.} This establishes a new heterogeneous paradigm where the pipeline is jointly constrained by GPU throughput and CPU memory efficiency.

\uline{Time Decomposition of KG Retrieval.}
Time breakdown in Fig.~\ref{fig:rag_macro_breakdown}(c) pinpoints the CPU bottleneck:  with large knowledge bases, Edge and Vertex Matching dominate retrieval time, while graph traversal (Edge/Vertex Traversal) and text processing (Get Context, Context Process) are minimal. This shows the primary cost is similarity matching computation over the expanding vector store, not complex graph operations. As the knowledge base grows, the linear-scaling vector store drives computation overhead, while sparse graph connectivity limits traversal growth. This divergence leads to a crucial architectural insight: \insight{\label{ins:rag_graph_vector_similarity} Graph-RAG workloads are ultimately constrained by similarity computations rather than graph traversal: similarity matching dominates CPU cost, while graph operations contribute minimally to overall latency.}

%To precisely identify the sources of CPU-side bottlenecks, we decompose the KG retrieval stage into six functional sub-stages: Edge Embedding, Vertex Embedding, Edge Retrieval, Vertex Retrieval, Get Context, and Context Process. The embedding sub-stages compute vector similarities to identify graph retrieval targets, while the remaining sub-stages handle neighborhood expansion, relevant-text acquisition, and text-level preprocessing.

%Time breakdown in Fig.~\ref{fig:rag_macro_breakdown}(c) reveals a striking pattern: with large knowledge bases, Edge Embedding and Vertex Embedding collectively dominate KG retrieval time, while graph traversal operations (Edge/Vertex Retrieval) and text processing (Get Context, Context Process) contribute minimally. This demonstrates that complex graph navigation and text manipulation are not the primary bottlenecks; instead, similarity computation over the expanding vector store accounts for the majority of CPU cost. As knowledge bases grow from 1.1 GB to 18 GB, the vector store scales linearly, driving a corresponding increase in similarity computation overhead. Graph structure traversal, in contrast, exhibits more modest growth due to the sparse connectivity typical of knowledge graphs. This divergence leads to a crucial architectural insight: \insight{\label{ins:rag_graph_vector_similarity} Graph-RAG workloads are ultimately constrained by vector operations rather than graph traversal: similarity computations dominate CPU cost, while structural graph operations contribute minimally to overall latency.}

\begin{figure}[!t]
\vspace{-10pt}

\centering
\includegraphics[width=0.95\textwidth]{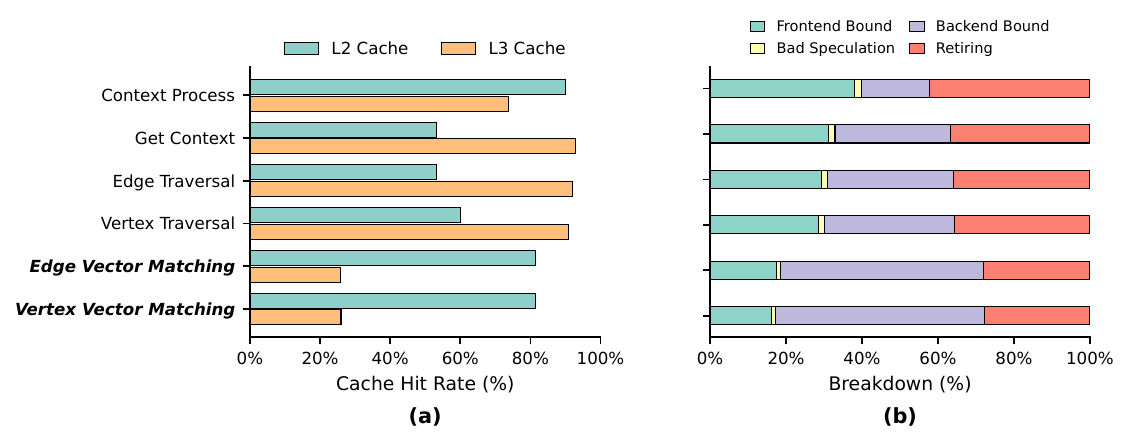}
\vspace{-15pt}
\caption{CPU microarchitectural analysis of retrieval stage (18 GB knowledge base): (a) Cache hierarchy efficiency across sub-stages; (b) Pipeline utilization patterns revealing bottleneck sources.}
\label{fig:rag_micro_breakdown}
\vspace{-15pt}
\end{figure}

\uline{Microarchitectural Characteristics of KG Retrieval.} We now examine the microarchitectural manifestations of KG retrieval bottlenecks. Cache analysis (Fig.~\ref{fig:rag_micro_breakdown}(a)) shows the dominant matching operations exhibit high L2 hit rates but low L3 utilization, characteristic of a streaming workload with strong spatial but weak temporal locality. Pipeline analysis (Fig.~\ref{fig:rag_micro_breakdown}(b)) further reveals these operations are predominantly Backend Bound, indicating CPU cores stall waiting for data despite efficient control flow. Combining these perspectives yields: \insight{\label{ins:rag_memory_bound} Similarity matching in Graph-RAG is a memory-bandwidth-bound streaming workload, constrained by memory throughput, not compute capability.}

\section{Optimization Guidelines}
\label{sec:summary_guidelines}

The key insights are summarized in Table~\ref{tab:key_insights}.

\begin{table}[t]
\vspace{-10pt}
\small
\centering
\caption{Summary of key insights.}
\label{tab:key_insights}
\vspace{-10pt}
%\resizebox{1.0\textwidth}{!}{%
\begin{tabular}{|p{13.5cm}|}
\hline
\textbf{(1) Performance Bottlenecks are Phase-Dependent and Dynamic} \\ \hline
%- The Prefill and Decode phases are bottlenecked by different resources: Prefill by computational throughput and Decode by memory bandwidth. \\
- A clear crossover point exists in the latency breakdown between Prefill and Decode as input length increases: Decode dominates at shorter inputs, while Prefill becomes the bottleneck beyond a critical length.
- At the operator level, the bottleneck migrates between the Feed-Forward Network and Attention. Feed Forward Network typically dominates at typical context lengths, whereas Attention becomes the primary bottleneck under extremely long contexts due to its quadratic (Prefill) or linear (Decode) complexity scaling. \\
- Microarchitecturally, Prefill is fundamentally compute-bound, with high Arithmetic Intensity and Tensor Core utilization, whereas Decode is memory-bound, constrained by frequent KV-Cache accesses and low Arithmetic Intensity. \\
\hline
\textbf{(2) Energy Consumption is Predictable and Decode-Centric} \\
- Total energy is dominated by the Decode phase and exhibits a strong linear relationship with the number of output tokens ($R^2 > 0.99$). The Prefill phase contributes negligibly to total energy. \\
- Energy consumption scales near-linearly with model parameter count, as larger models incur both higher instantaneous power draw and longer execution times. \\
- Combined with the predictability of output length from prompts, energy consumption can be accurately forecasted before request execution. \\
\hline
\textbf{(3) Scaling Laws are Governed by Phase Characteristics} \\ \hline
- In multi-GPU setups, Prefill benefits from Tensor Parallelism due to its compute-bound nature, achieving near-linear scaling. In contrast, Decode gains little from Tensor Parallelism and is often optimal on a single GPU or with Pipeline Parallelism, as communication overhead outweighs computational benefits. \\
- On resource-constrained edge devices, constraints are amplified: memory bandwidth limitations exacerbate Decode bottlenecks, while thermal/power constraints trigger frequency throttling that particularly impacts Prefill performance. \\
\hline
\textbf{(4) Emerging Architectures and Workflows Reshape Bottlenecks} \\ \hline
- MoE decouples model capacity from computational cost, achieving superior latency and throughput by activating a subset of parameters. However, they introduce significant routing overhead in Decode, which becomes a major performance factor. \\
- In RAG workflows, scaling the knowledge base triggers a bottleneck shift from GPU computation to CPU-side memory access during retrieval operations, forming a heterogeneous CPU-GPU execution paradigm. \\
\hline
\end{tabular}%
%}
\vspace{-20pt}
\end{table}

Grounded in our systematic characterization and insights, we propose the following guidelines.

\textbf{System Architecture and Scheduling.} 
\uline{(1) Phase-Decoupled Serving.} 
Route Prefill to compute-optimized instances and Decode to bandwidth-rich hardware (\insightref{ins:phase_resource_bottleneck}, \insightref{ins:compute_vs_memory_bound})~\cite{Patel2024Splitwise,Zhong2024DistServe,Hu2025ShuffleInfer}. Use Chunked Prefill for interleaved execution where hardware disaggregation is infeasible, mitigating the \emph{latency trap} in mixed workloads (\insightref{ins:throughput_latency_tradeoff})~\cite{Agrawal2024Sarathi}.
\uline{(2) Predictability-Driven Scheduling.}
Employ Shortest Remaining Processing Time scheduling based on linear Prefill latency scaling with uncached tokens (\insightref{ins:prefill_linear_uncached})~\cite{du2025prefillonly}. Implement energy-aware throttling and placement strategies using the predictable linear relationship between energy consumption and output length (\insightref{ins:energy_linear_output}, \insightref{ins:energy_prompt_predictable}).

\textbf{Microarchitecture and Kernel Optimization.}
\uline{(1) Prefill Latency Hiding.} 
Address Tensor Core saturation and HMMA instruction latency through instruction-level parallelism, software pipelining, and K-tile double buffering~\cite{Hong2024FlashDecodingPP} (\insightref{ins:compute_vs_memory_bound}, \insightref{ins:prefill_dual_saturation}, \insightref{ins:roofline_phase_heterogeneity}).  Utilize persistent GEMM-style kernels and warp-level prefetching to sustain pipeline occupancy~\cite{Dao2022FlashAttention}.
\uline{(2) Decode Weight Optimization.}
Apply weight-only quantization to memory-bound FFN-Up layers to alleviate bandwidth pressure~\cite{Lin2024AWQ,Frantar2023GPTQ}(\insightref{ins:decode_gemm_heterogeneous}, \insightref{ins:decode_ffn_weight_bandwidth}). For execution-bound O-Proj/FFN-Down kernels, optimize register allocation or employ operator fusion (\insightref{ins:decode_gemm_heterogeneous}).
\uline{(3) KV-Cache Bandwidth Management.}
Reduce memory footprint via Grouped Query Attention~\cite{Ainslie2023GQA} and KV-cache quantization~\cite{Hooper2024KVQuant} (\insightref{ins:decode_dram_locality_shift}, \insightref{ins:attncore_locality_degradation}, \insightref{ins:long_context_attention_bottleneck}). 
%Implement paging and eviction-aware policies to minimize fragmentation.

\textbf{Parallelism Strategy Selection.}
\uline{(1) Phase-Adaptive Parallelism.}
Assign Tensor Parallelism to Prefill-heavy workloads and Pipeline Parallelism (or single-GPU execution) to Decode-heavy scenarios(\insightref{ins:decode_single_gpu_better}, \insightref{ins:prefill_tp_scaling}, and \insightref{ins:phase_aware_parallelism} ). Employ dynamic parallelism switching or heterogeneous configurations.
\uline{(2) Communication-Computation Overlap.}
Split projection matrices into moderate-sized chunks to enable overlapping of collective communication with GEMM computation (\insightref{ins:decode_single_gpu_better}, \insightref{ins:kernel_comm_balance}). 

\textbf{Domain-Specific Optimizations.}
\uline{(1) Bandwidth-Aware Edge Execution.}
Prioritize bandwidth conservation and thermal stability over peak throughput (\insightref{ins:jetson_kv_memory_limit}, \insightref{ins:edge_ffn_memory_bound}, \insightref{ins:edge_dvfs_heterogeneity}).  Limit batch sizes and optimize memory layouts for LPDDR characteristics (\insightref{ins:edge_dvfs_heterogeneity}).
\uline{(2) Locality-Aware MoE Routing.}
Implement expert-affinity scheduling to improve weight locality under small effective batches (\insightref{ins:moe_decode_overhead_routing}, \insightref{ins:moe_l2_locality_batch_sensitive}). Group requests activating the same experts while balancing load  (\insightref{ins:moe_small_effective_batch}).
\uline{(3) Accelerated Retrieval in RAG.}
Offload similarity computation to near-memory accelerators when feasible (\insightref{ins:rag_kg_memory_constraint}, \insightref{ins:rag_memory_bound}).

\section{Conclusion}
\label{conclusion}

This study establishes a systematic multi-level characterization for LLM inference optimization, encompassing microarchitectural profiling, energy consumption analysis, and system scaling evaluation. The resulting insights provide both foundational principles for architecture-system co-design and actionable optimization strategies, enabling efficient LLM deployment across heterogeneous computing environments spanning from cloud data centers to resource-constrained edge devices.

\bibliographystyle{ACM-Reference-Format}
\bibliography{references.bib}

\end{document}